\documentclass[compsoc, conference, a4paper, 10pt, times]{IEEEtran}

\usepackage{cite}
\usepackage{amsmath,amssymb,amsfonts}
\usepackage{algorithmic}
\usepackage[pdftex]{graphicx}
\usepackage{textcomp}
\usepackage{xcolor}
\usepackage{booktabs}
\usepackage[hidelinks]{hyperref}

\usepackage{inconsolata}
\usepackage{url}
\usepackage{tikz}
\usepackage{amsmath,amsfonts,amssymb,amsthm}
\usepackage{booktabs}
\usepackage[libertine]{newtxmath}
\usepackage{cite}
\usepackage{babel}

\usepackage{xcolor}
\usepackage{mathtools}
\usepackage{bm} 
\usepackage[inline]{enumitem}
\usepackage{harveyballs}
\usepackage{xspace}

\usepackage{listings}
\usepackage{lstlangarm}
\definecolor{light-gray}{gray}{0.95}

\lstdefinestyle{C-extended}{
language=C,
basicstyle=\ttfamily\footnotesize,
breakatwhitespace=false,
mathescape=true,
captionpos=b,
numbers=left,
numbersep=5pt,
numberstyle=\tiny\color{gray},
otherkeywords={uint8_t,uint16_t,uint32_t,uint64_t}, 
emph={__delay_ms, defined, true, false}, 
}

\newcommand{\red}[1]{\color{red!70!black}{#1}\color{black}}
\newcommand{\green}[1]{\color{darkgreen}{#1}\color{black}}
\newcommand{\lightgreen}[1]{\color{darkgreen}{#1}\color{black}}
\newcommand{\gray}[1]{\color{gray}{#1}\color{black}}

\newcommand{\B}[1]{\ensuremath{\text{P}_\text{{#1}}}} 

\makeatletter
\let\old@lstKV@SwitchCases\lstKV@SwitchCases
\def\lstKV@SwitchCases#1#2#3{}
\makeatother
\usepackage{lstlinebgrd}
\makeatletter
\let\lstKV@SwitchCases\old@lstKV@SwitchCases

\lst@Key{numbers}{none}{%
    \def\lst@PlaceNumber{\lst@linebgrd}%
    \lstKV@SwitchCases{#1}%
    {none:\\%
        left:\def\lst@PlaceNumber{\llap{\normalfont
                \lst@numberstyle{\thelstnumber}\kern\lst@numbersep}\lst@linebgrd}\\%
        right:\def\lst@PlaceNumber{\rlap{\normalfont
                \kern\linewidth \kern\lst@numbersep
                \lst@numberstyle{\thelstnumber}}\lst@linebgrd}%
    }{\PackageError{Listings}{Numbers #1 unknown}\@ehc}}
\makeatother

\usepackage{soul}
\usepackage{pdflscape}
\usepackage{nameref}
\usepackage{subcaption}

\usepackage{makecell}
\usepackage{tabularx,ragged2e}
\usepackage{multirow}
\usepackage[flushleft]{threeparttable}

\newcommand{\tool}{Pascal\xspace}

\newcolumntype{g}{>{\columncolor{babyblueeyes!25}}c}
\newcolumntype{C}{>{\centering\arraybackslash}X}
\newcolumntype{R}{>{\raggedleft\arraybackslash}X}
\newcolumntype{L}{>{\raggedright\arraybackslash}X}

\definecolor{darkgreen}{rgb}{0.0, 0.5, 0.0}
\definecolor{babyblueeyes}{rgb}{0.63, 0.79, 0.95}

\usepackage{pifont}
\newcommand{\gcmark}{\textcolor{darkgreen}{\ding{51}}}
\newcommand{\cmark}{\ding{51}}

\newcommand{\xmark}{{\ding{55}}}
\usepackage[frozencache=true,cachedir=minted-cache]{minted} 

\theoremstyle{definition}
\newtheorem{definition}{Definition}
\newtheorem{observation}{Observation}

\definecolor{babyblue}{rgb}{0.54, 0.81, 0.94}
\definecolor{airforceblue}{rgb}{0.36, 0.54, 0.66}
\definecolor{airforceblue1}{rgb}{0.36, 0.70, 0.90}
\definecolor{lightskyblue}{rgb}{0.53, 0.81, 0.98}
\definecolor{tealgreen}{rgb}{0.0, 0.51, 0.5}
\definecolor{azure}{rgb}{0.94, 1.0, 1.0}
\long\def\com#1{}

\DeclareRobustCommand\circled[1]{\tikz[baseline=-0.6ex]{\node[shape=circle,fill,inner sep=1.2pt] (char) {\textcolor{white}{\scriptsize\bf#1}}}}

\usepackage[ruled,linesnumbered, vlined]{algorithm2e}
\usepackage{etoolbox}
\makeatletter
\patchcmd{\@algocf@start}
{-1.5em}
{0pt}
{}{}
\makeatother

\usepackage{hhline}
\usepackage{comment}

\graphicspath{ {./figures/} }
\long\def\ferhat#1{{#1}}
\long\def\rebuttal#1{}
\long\def\revision#1{#1}

\newif\ifccstemplate
\ccstemplatefalse

\begin{document}

\title{Towards Automated Detection of Single-Trace Side-Channel Vulnerabilities in Constant-Time Cryptographic Code}


\author{\IEEEauthorblockN{Ferhat Erata}
\IEEEauthorblockA{\textit{Yale University} \\
}
\and
\IEEEauthorblockN{Ruzica Piskac}
\IEEEauthorblockA{\textit{Yale Unviersity} \\
}
\and
\IEEEauthorblockN{Victor Mateu}
\IEEEauthorblockA{\textit{Technology Innovation Institute} \\
}
\and
\IEEEauthorblockN{Jakub Szefer}
\IEEEauthorblockA{\textit{Yale Unviersity} \\
}
}

\maketitle

\begin{abstract}
    Although cryptographic algorithms may be mathematically secure, it is often possible to leak secret information from the implementation of the algorithms. Timing and power side-channel vulnerabilities are some of the most widely considered threats to cryptographic algorithm implementations. Timing vulnerabilities may be easier to detect and exploit, and all high-quality cryptographic code today should be written in constant-time style. However, this does not prevent power side-channels from existing.  With constant time code, potential attackers can resort to power side-channel attacks to try leaking secrets. Detecting potential power side-channel vulnerabilities is a tedious task, as it requires analyzing code at the assembly level and needs reasoning about which instructions could be leaking information based on their operands and their values. To help make the process of detecting potential power side-channel vulnerabilities easier for cryptographers, this work presents Pascal: Power Analysis Side Channel Attack Locator, a tool that introduces novel symbolic register analysis techniques for binary analysis of constant-time cryptographic algorithms, and verifies locations of potential power side-channel vulnerabilities with high precision. Pascal is evaluated on a number of implementations of post-quantum cryptographic algorithms, and it is able to find dozens of previously reported single-trace power side-channel vulnerabilities in these algorithms, all in an automated manner. 
\end{abstract}


\section{Introduction}
\label{introduction}

Although cryptographic algorithms may be formally proven to be secure, often it is the case that there is a gap between a mathematical formalization of the algorithm and its implementation; the implementation might leak the secret information. When one can infer secret information from either observing the running time of the program~\cite{aciicmez2005improving, bernstein2005cache, brumley2011remote, yarom2017cachebleed}, or its power consumption~\cite{DPA, kocher2011introduction, mangard2008power}, or electromagnetic emanations (EM)~\cite{quisquater2001electromagnetic}, traditionally this is referred to as a side-channel vulnerability. Timing side-channel vulnerabilities have been studied the most, and currently we have numerous powerful analysis techniques and tool that can detect
if code is resistant against timing side-channel attacks or to help write so-called constant-time code~\cite{daniel2020binsec, ct2019verifying, brennan2018symbolic, wichelmann2018microwalk, he2020ct,  antonopoulos2017decomposition, athanasiou2018sidetrail, wu2018eliminating, borrello2021constantine, chattopadhyay2018symbolic}.

However, even if the programmers follow the constant-time paradigm, this still does not prevent power side-channels attacks. Computing and manipulating different data
causes different power consumption. An attacker with physical access to sampling power consumption~\cite{hw-hack-book, 2014:204}, or with remote access to power-related performance counters~\cite{platypus}, can observe power variations as an algorithm executes. This can leak secure data.

If the code is non-constant time, the attackers already have an easier (timing) attack that they
can leverage. Therefore, in this paper, we consider only constant-time code.
The power side-channels that still remains in constant-time code are often difficult to detect, even for experts.
Therefore, the focus of our work is on developing automated reasoning methods to find potential locations of power side-channel leaks in cryptographic code. The existing work mostly relies on hypothesis testing~\cite{tvla,2015:207,2013:298,sost}  and
known power side-channel attacks are found experimentally by collecting code traces and conducting various statistical analyzes or using target specific leakage simulators~\cite{elmo:1, rosita, emsim}.
Without collecting traces, our approach analyzes the binaries and pinpoint location(s) of potential single-trace power side-channel vulnerabilities, which can then be confirmed by using methods such as TVLA, and then finally fixed. We built a tool, \tool, and empirically evaluated it on $30$ publicly disclosed power-side channel vulnerabilities mainly in post-quantum cryptographic algorithms; \tool found all previously reported single-trace vulnerabilities in these algorithms.

\tool combines differential program analysis~\cite{differentialProgramAnalysis, farina2019relational,shadowSymbolicExecution} with optimization queries~\cite{z3opt, nadel2016bit, symba,OptiMathSAT} in order to identify the most vulnerable locations in the binaries of a constant-time code.
We use Hamming weight and Hamming distance leakage models~\cite{kocher2011introduction, DPA, CPA}. The Hamming weight, defined on a binary string, is the number of 1's in the string. The Hamming weight leakage model assumes that the Hamming weight of the operands is strongly correlated with the power consumption.

In symbolic analysis, some or all variables (or in the context of binary analysis, registers or memory locations) are represented by a symbolical variable. This symbolical variable represents all possible
values that the variable could take.
As the execution proceeds over each instruction, symbolic execution derives logical formulas defined using these symbolical variables. These formulas, defined in symbolical variables,  represent a summary of a program execution. We use these formulas to derive limits for the
range of values the symbolical variable can represent.
In differential part of the analysis, we create optimization queries to find the minimum and maximum Hamming weight differences for operands for each of the instructions. At the binary code level, if the Hamming weight difference is typically large in destination registers, large power measurement difference is assumed, and the code location can be vulnerable to practical power analysis attacks in that the secret can be guessed with one or few traces.

Our tool \tool, at high-level, formally locates these vulnerable instructions in the binaries where there is a large separation between the Hamming weights of \emph{register writes}. For example, through the symbolic execution and the optimization queries the tool may find that some instruction only can have secret-related inputs of either $0x00\ldots 0$ or $0xFF\ldots F$, then this instruction may lead to potential power side-channel vulnerability since power consumption of computation related to $0x00\ldots 0$ is much different from for $0xFF\ldots F$. Automatically analyzing the code to find all the possible ranges of register values, and their Hamming weights, is a non-trivial task and one of the main contributions of this~paper.

\begin{figure}[t]
    \centering
    \includegraphics[trim={0.02cm 0.01cm 0.5cm 0.01cm}, clip, width=\linewidth]{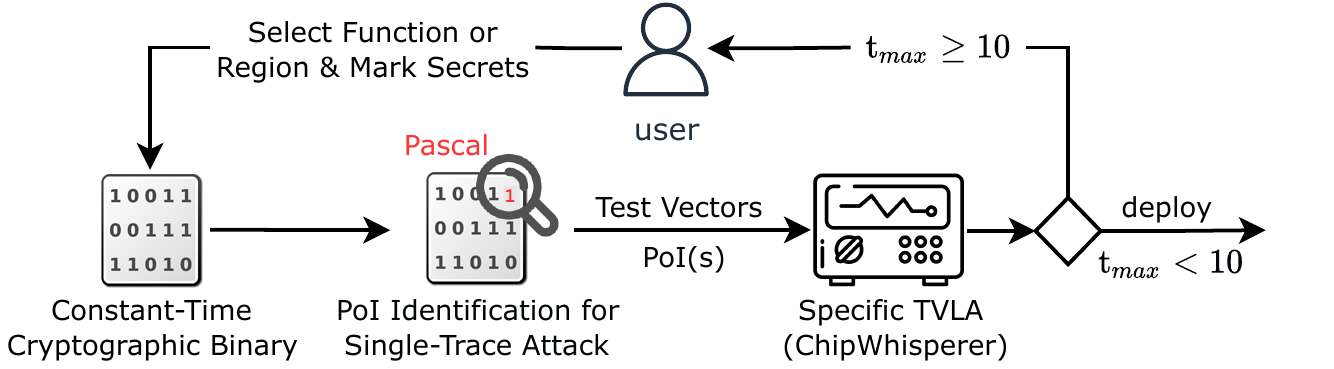}
    \caption{\tool reports addresses of Point of Interests (PoI) (vulnerable instructions) in the binary and generates Test Vectors for TVLA analysis \revision{if wanted.}~\rebuttal{\#3}}\label{fig:method}
\end{figure}

The methodology of \tool is shown in Figure~\ref{fig:method}.
\revision{
     \tool is designed to be used in a standalone manner, allowing developers to easily check for single-trace vulnerabilities in their code. However, once a point of interest is found, developers can then use TVLA or other statistical analysis methods to confirm the vulnerability empirically. To aid in this process, Pascal also generates test vectors. 
}~\rebuttal{\#3}

With our tool, we can analyze cryptographic code to get the differential Hamming weights of all the possible operands for all instructions in a target function. The input code can be in a variety of programming languages, as it is first compiled into binary before analysis is done.
\tool~can naturally explore impacts of different target architectures and optimizations by compiling the code with the different flags or optimization levels and then running the analysis on the different binaries.

To empirically evaluate \tool, we examined, to the best of our knowledge, all known power single-trace side-channel attacks against the publicly available constant-time implementations of post-quantum cryptographic algorithms submitted to NIST's post-quantum cryptography standardization process. Our method successfully pinpointed more than 30 known power-side channel vulnerabilities in constant-time implementations in a variety of cryptographic schemes, from Elliptic Curve~\cite{ecc-based} to Lattice-based~\cite{lattice-based} and to Lightweight~\cite{2021:nist} cryptography. \tool is the first tool that can find the locations of potential vulnerabilities in an automated way with limited programmer input: they only need to specify a function of interest and which secret inputs are related to it, and the tool finds if and where there are instructions that could be sources of power side-channel leakage.

Detecting potential power side-channel vulnerabilities is a tedious task, as they require analyzing code at the {\itshape assembly level} and need reasoning about which instructions could be leaking information based on their operands. Our tool is designed to automatize this procedure by {\itshape formally} analyzing given binaries. It is developed on top of \texttt{angr}~\cite{angr,driller,firmalice} binary analysis framework, \texttt{Z3}~\cite{z3} and \ferhat{\texttt{CVC5}~\cite{cvc5} SMT solvers, and single-objective linear optimization algorithms over bitvector terms~\cite{z3opt, symba, OptiMathSAT:CAV} of \texttt{Z3}~\cite{nuz} and \texttt{OptiMathSAT}~\cite{OptiMathSAT} solvers}.

\revision{
    Our work is one of the first to approach the problem of detecting power side-channel vulnerabilities using a {\em white-box} technique that considers the semantics of the tainted instruction flow and the symbolic states of the registers. Our work is also the first to adapt the relatively new concept of {\em Relational Symbolic Execution}~\cite{farina2019relational, differentialProgramAnalysis} for power side-channel analysis.
}~\rebuttal{\#2}

\subsubsection*{Contributions}
In summary, this work makes the following contributions:
\begin{enumerate*}[nosep, label=\textbf{(\arabic*)}]
    \item Definition of a robustness measure of cryptographic code against power side-channel attacks, which reveals and characterizes the Hamming weight patterns of instruction operands. From an attacker's perspective, it allows them to identify the most vulnerable code regions for an attack, whereas, from a cryptographic library developer's perspective, allows them to pinpoint the most vulnerable code regions for hardening them~(see {\S\ref{sec:overview}}).
    \item Development of the novel prototype analysis tool for automatic detection of potential power side-channel vulnerabilities in constant-time implementations of cryptographic algorithms at the binary code level~(see {\S\ref{sec:approach}}).
    \item Systematic analysis of the power side-channel attacks known in the literature to validate
          that \tool can identify the most vulnerable binary code locations of a constant-time code under Hamming weight or Hamming distance leakage model. We successfully identified $30$ different vulnerabilities in constant-time implementations in a variety of cryptographic schemes~(see {\S\ref{sec:vulnerabilities}}).
\end{enumerate*}

\subsubsection*{Availability} \url{https://github.com/ferhaterata/pascal}

\begin{figure*}[!th]
  \centering
  \begin{minipage}[t]{.97\columnwidth}
    \ifccstemplate\else\small\fi
    \setlength{\tabcolsep}{0.5em}
    \begin{tabular}{@{}lcl@{}}
      constant-time code &                    & {lifted binary code}
      \\
      \midrule
      $
        \begin{aligned}
          f(sum_{[8]S} & ,\; \red{x}_{[8]S}):               \\
          \red{mask}   & := (\red{x} - 64) \gg 7            \\
          \red{sum}    & := sum + (\sim mask \land \red{x}) \\
        \end{aligned}
      $
                         & $\rightsquigarrow$ &
      $
        \begin{aligned}
           & f({sum_0}_{[8]S}, \, \red{x_0}_{[8]S}):      \\
           & \quad\red{r_0}:= \red{x_0} - 64              \\
           & \quad\red{r_1}:= \red{r_0} \gg 7 \qquad\star \\
           & \quad\red{r_2}:= \, \sim \red{r_1}           \\
           & \quad\red{r_3}:= \red{r_2} \land \red{x_0}   \\
           & \quad\red{r_4}:= sum_0 + \red{r_3}           \\
        \end{aligned}
      $
    \end{tabular}
    \caption{Example of conditional addition written in a constant-time style using masking.}\label{fig:csubq}
  \end{minipage}
  \qquad
  \begin{minipage}[t]{.97\columnwidth}
    \ifccstemplate
      \lstdefinestyle{mwe}{basicstyle=\ttfamily\small}
    \else
      \lstdefinestyle{mwe}{basicstyle=\ttfamily\footnotesize}
    \fi
    \ifccstemplate\else\small\fi
    \begin{tabular}{@{}p{0.45\columnwidth}|p{0.5\columnwidth}@{}}
      Arm32, O0                                                                                                                                   & O3, cortex-m4, armv7e-m \\
      \midrule
      \begin{minipage}[t]{.45\linewidth}
\begin{lstlisting}[style=mwe, language={[arm]Assembler}, numbers=none]
mov r3, r0       
...
sub r3, r3, 0x40
asr r3, r3, 7     $\red{\star}$
...
bx lr
\end{lstlisting}\end{minipage} &
      \begin{minipage}[t]{.45\linewidth}
\begin{lstlisting}[style=mwe, language={[arm]Assembler}, numbers=none]
sub.w r3, r1, 0x40     
bic.w r1, r1, r3, asr 7 $\red{\star}$
add r0, r1             
uxtb r0, r0
bx lr
\end{lstlisting}\end{minipage}
    \end{tabular}
    \caption{Disassembly of conditional addition from Figure~\ref{fig:csubq} based on different compiler flags. }\label{fig:dis-cadd}
  \end{minipage}
\end{figure*}

\section{Motivating Example}
In this section, we illustrate a power side-channel leakage in a constant-time code and its detection.

\subsubsection*{A Constant-Time Code}\label{sec:motivating-example}
It is a well-known fact that the power consumption during certain stages of a cryptographic algorithm exhibits a strong correlation with the Hamming weight of its underlying variables, i.e., Hamming weight leakage model~\cite{platypus, CPA, 2005:338, mangard2008power, 2010:180, mdpi:2020}.
This phenomenon has been widely exploited in the cryptographic literature in various attacks targeting a broad range of schemes~\cite{2021:HOST, 2021:1307, 2020:1559, 2020:992, 2020:368, 2020:549, 2019:335, 2019:1236, 2021:790, 2021:874, 2018:1, 2019:948, 2021:858, 2016:923, 2021:104, 2021:101, sim2017key, sim2018key}. Due to the intrinsic connection between the Hamming weight of intermediate cipher variables and the power consumption of software implementations of cryptographic algorithms, the most vulnerable binary locations of a constant-time code are those that are most likely to be executed on a set of few intermediate values whose Hamming weight differences are considerably large.
For example, consider the conditional addition function, $if (\red{x}\geqslant 64) \;then \;\red{sum} := sum + \red{x}$, where $sum$ is the running sum and $x$ is the next secret byte. In this simple example, the tainted intermediate variables related to the secret input $x$ are colored red to help visualize the propagation of the secret information (aka. forward tainting). This code has obvious timing vulnerability, since execution of the addition operation depends on the secret value $x$. However, this code can be converted to constant-time code using {\itshape masking}, as shown in Figure~\ref{fig:csubq}.

\begin{figure}
  \includegraphics[trim={0.66cm 0.15cm 1.5cm 0.3cm}, clip, width=\linewidth]{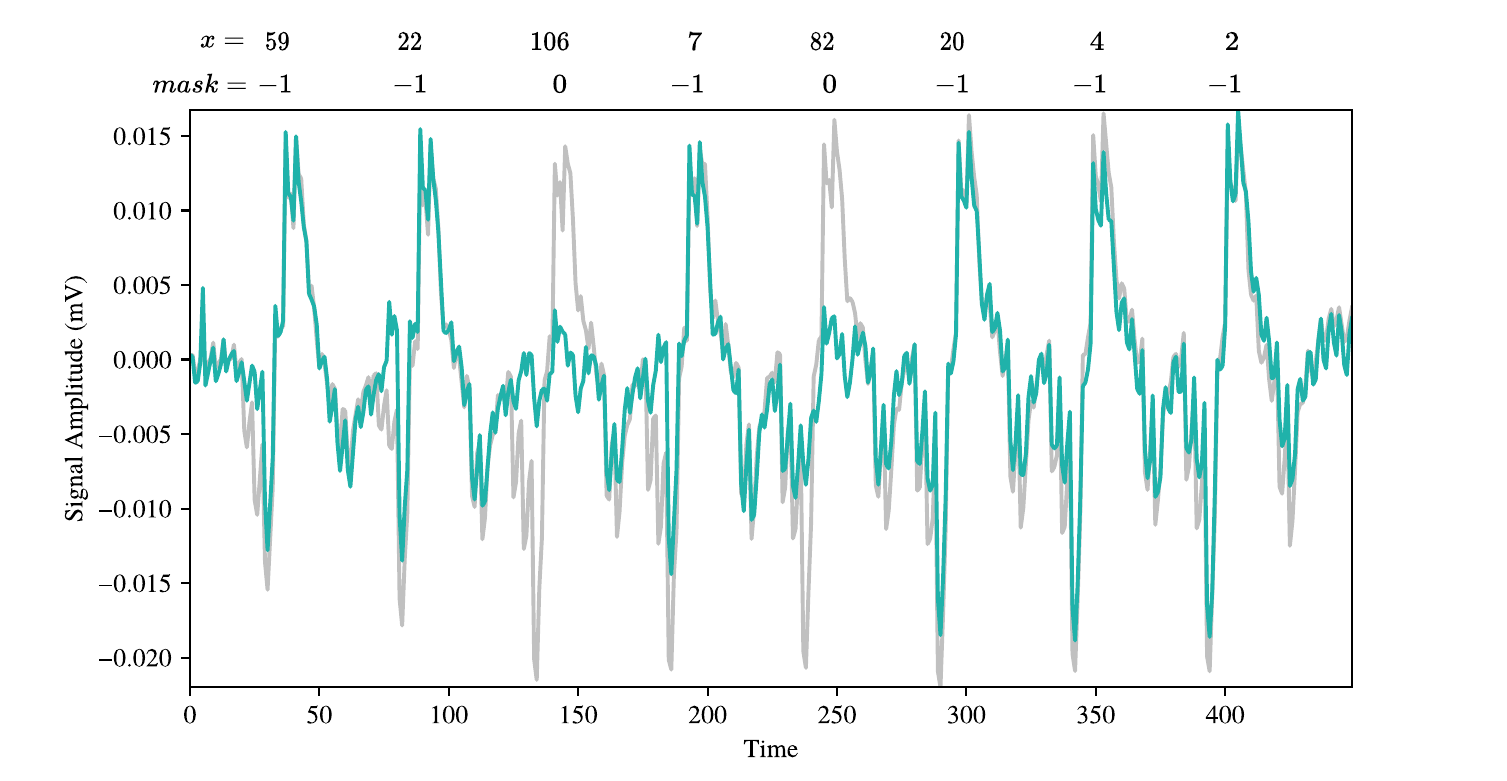}
  \caption{A {\itshape single-trace template-based attack}. The trace in green exposes the differences between the mask value being either 0 or -1. The reference trace in gray is a template having the max. \textit{Hamming weight difference}.}
  \label{fig:diff-power-analysis}
\end{figure}
In the constant-time version of the conditional addition function, a {\itshape mask}, containing 0 or -1 ($= 0xFF$), replaces the if-condition. The mask calculation is shown in a simplified machine code in Figure~\ref{fig:csubq}. An arithmetic right shift operation (indicated by a $\star$) extracts the sign bit of the 8-bit intermediate value, $m_{0{[8]S}}$ ($S$ means signed two's complement). The processed message bit is leaked
\begin{enumerate*}[label=\textbf{(\roman*)}]
  \item neither in a branch,
  \item nor in an address-index look-up,
  \item nor through a loop bound,
  \item nor in differences in execution time,
\end{enumerate*}
but through power consumption which differs between processing operand of all zeros vs. all ones; and here the {\itshape mask} either contains exactly either ones or zeroes only. Chances that these two different processed values can be detected by analyzing the power consumption of the device are very high. This will give attackers a good attack point (aka. Point of Interest (PoI) or Point of Leakage) for the analysis of the power consumption of a constant-time code under hamming weight leakage model,
since in particular an attacker will infer a statistical property of the secret value, which is in our case to be either in the range of $[-128, 63]$ or $[64, 127]$.
The PoIs are identified in the binaries of the code.

In Figure~\ref{fig:dis-cadd}, the binary outputs of \texttt{arm-none-eabi-gcc} compiler toolchain targeting 32-bit Arm architecture with the different compiler flags generates two different vulnerable locations. \texttt{asr} instruction performs arithmetic shift right operation over operands; and \texttt{bic.w}, Bitwise Bit Clear, performs a bitwise AND of a register value and the complement of an immediate value, and writes the result to the destination register. As we can see that different compiler flags generate different code and different vulnerable locations, it is necessary to focus on the assembly code level for the power side-channel checking.

\subsubsection*{Test Vector Leakage Assessment}
The Test Vector leakage Assessment (TVLA)~\cite{tvla} identifies differences between two sets of side channel measurements by computing the t-test for the two sets of measurements. It is being used in the literature to confirm the {\itshape presence} or {\itshape absence} of side leakages for power traces~\cite{mdpi:2020, 2015:207, 2013:298, 2017:138, 2015:1215, 2017:624}.

To empirically show that the vulnerability exists in the conditional addition function, we perform a {\itshape single-trace template-based attack} using differential power analysis~\cite{DPA,template} 
on ChipWhisperer UFO STM32F3 target board~\cite{2014:204}. STM32F303 is a microcontroller based on the 32-bit ARM Cortex-M4 processor core, which is commonly used in embedded systems such as IoT/Edge devices. 
Based on the {\itshape Point of Interest} that we identified at line 2 in Figure~\ref{fig:csubq}, we fix the $sum = 0$ and randomly draw eight sensitive $s$ values from two sets, $[-128, 63]$ or $[64, 127]$, to take conditionally running sum of those eight values. We repeat this procedure to obtain 100 traces from each.
In Figure~\ref{fig:diff-power-analysis}, we show averaged differential power traces of the two sets to create the {\itshape template} in gray color. The green one is a {\itshape single-trace}, having obtained by triggering the function $f$ eight times.
Evidently, the difference between a mask value being 0 or -1 is immediately visible in the respective trace when $f$ is compiled at -O3.

\section{Overview}
\label{sec:overview}

Section \S\ref{sec:power-analysis} explains the data-dependent power consumption phenomenon in CMOS circuits, and Section \S\ref{sec:hamming-weight-leakage-model} formalizes the Hamming weight and Hamming distance leakage models. Section \S\ref{sec:diff-power-analysis} introduces the notion of differential Hamming weights and Hamming distances and relates it to {differential program analysis}. Finally, in Section \S\ref{sec:entropy}, we introduce $\omega$-class sampling model to quantify self-information content based on possible Hamming weight classes at a register.

\subsection{Data Dependent Power Consumption}
\label{sec:power-analysis}

Power analysis attacks are built upon the observation that the power consumption of CMOS digital circuits is data-dependent \textit{by design}. Each bit flip requires one or more voltage transitions from 0 to high (or vice versa). Different data values typically entail differing numbers of bit flips and therefore produce distinct power traces~\cite{power-analysis}.
Therefore, any circuit not explicitly designed to be resistant to power attacks has data-dependent power consumption. However, in a complex circuit, the differences can be so slight that they are difficult to distinguish from a \textit{single trace}, particularly if an attacker's sampling rate is limited~\cite{platypus,hertzbleed}.
Therefore, it is necessary to use statistical techniques such as Differential Power Analysis~\cite{DPA} and Correlation Power Analysis~\cite{CPA} across multiple power traces~\cite{platypus}.
\ferhat{In these analysis methods the \emph{Hamming weight} or \emph{Hamming distance} leakage models are preferred to quantify the power consumption of a CMOS circuit.}

\subsection{Hamming Weight and Hamming Distance}
\label{sec:hamming-weight-leakage-model}

\subsubsection*{Leakage Models} In the value-based leakage model, the leakage correlates with the Hamming Weight of values at registers leaked by instruction execution, that differs from the transition-based model where the value leaked by an instruction correlates with the Hamming distance between the result of its new value and the previous old value in the register. \tool~supports both models. Additionally, \tool incorporates an information theoretical, value-based measure, namely Hamming weight class sampling model, to quantify leakages.

\begin{definition}[Hamming weight]\label{def:hw}
  Given an n-bit element $v \in \mathbb{F}_n$ , let $0 \le \omega(v) \le n$ be its Hamming weight, i.e., the number of bits that are set to one in the binary representation of $v$.
\end{definition}

The leakage model considered in this paper are the \emph{Hamming weight} \ferhat{and \emph{Hamming distance}} leakage models~\cite{kocher2011introduction, DPA, CPA}. We use $\omega(v)$ to refer to the Hamming weight of a value $v$ \ferhat{and use $d(v, v^\prime)$ to refer the Hamming distance of values $v$ and $v^\prime$}.
\begin{definition}[\ferhat{Hamming distance}]\label{def:hd}
  \ferhat{
    Hamming distance is a metric for measuring the edit distance between two sequences.
    Given two n-bit elements $v, v^\prime \in \mathbb{F}_n$, let $0 \le d(v, v^\prime) \le n$ be their Hamming distance, i.e., $d(v, v^\prime) \triangleq \omega(x \oplus y)$. $\uparrow d(v, v^\prime)$ and $\downarrow d(v, v^\prime)$ represents {\itshape maximum} and {\itshape minimum} Hamming distance between $v$, $v^\prime$, respectively.
  }
\end{definition}

Due to noise in power measurements, it is not possible to directly deduce the Hamming weight for a particular value of interest, that is being updated inside the central processing unit of the microcontroller or a microprocessor. It is nevertheless possible to find out the Hamming weight of a particular value by the means of a differential power analysis.

\begin{definition}[Power Consumption \& Power Trace]
  Given a function $F: \mathbb{F}_m \mapsto \mathbb{F}_n$ that executes a sequence of instructions for an arbitrary $x \in \mathbb{F}_m$, we denote by $P(F(x))$ the power consumption of this function and by $\overline{P}(F(x))$ its average over multiple repetitions. A power trace $P_{T_n}(F(x))$ is a vector of $n$ values of power samples, where $n$ is the number of samples taken at each time instant over the execution of function $F$.
\end{definition}

We experimentally verified that this property holds on a number of microcontrollers. In Figure~\ref{fig:hw-vs-delta-hw} \ferhat{on the left graph}, the function $F$ corresponds to execution of $R_d := R_d \gg x$ including an \texttt{asr} instruction.
Note that knowledge of $\omega(F(x))$ does not necessarily mean that $F(x)$ is uniquely recovered. However, $0$ and $2^n - 1$ within $\mathbb{F}_n$ are the only elements with Hamming weight 0 and $n$ respectively.

\begin{definition}[Monotonic relation between Power Consumption and Hamming weight]\label{def:monotone}
  If $\omega(F(x)) > \omega(F(y))$ then $\overline{P}(F(x)) > \overline{P}(F(y))$ for two different $x, y \in \mathbb{F}_n$.
\end{definition}

\revision{The power consumed by the attacked device has to be a monotonic function of the Hamming weight of the processed data for the Hamming weight leakage model to work. This observation is validated experimentally in the literature such as on Intel and ARM's mobile, desktop, and server CPUs~\cite{platypus}. We also confirmed that 8-bit ATXmega128D4-AU (AVR instruction set), 32-bit STM32F3/STM32F4 (ARM Cortex-M4), STM32F0 (ARM Cortex-M0) microcontrollers exhibits a monotonic relation between power consumption and Hamming weights of data, and we have also found that the energy consumption of ultra-low powered devices (MSP430 instruction set of the MSP430FR5994 microcontroller) is monotonic with respect to the Hamming weight of the processed data (STM32F3 in Figure~\ref{fig:hw-vs-delta-hw}). 
  Almost all the power and EM side-channel attack papers use the Hamming weight as a metric and accept monotonicty assumptions for the power consumption behavior of the attacked~device.}
~\rebuttal{\#1.b}

\subsection{Differential Power Analysis}\label{sec:diff-power-analysis}

Different code execution paths may leave different power traces, and this may lead to different instructions for each path and affect power consumption, this can be also classified as a timing vulnerability.
This occurs based-on secret dependent branches.
Since constant-time code always generates the same length of power traces, we omit this type of behavior in our analysis. In fact, there is no secret-dependent branching and secret-dependent loop bounds among the Post-Quantum Cryptographic (PQC) implementations~\cite{pqm4} nor in reference implementations among finalists of NIST's PQC standardization process~\cite{2020:nist:pqc} which we analyze later in this work. However, even with the same control flow, different data being manipulated in the microcontroller or the microprocessor affects power consumption.

\begin{definition}[Differential Hamming weight]\label{def:diff-hw}
  Given two n-bit elements $v, v^\prime \in \mathbb{F}_n$, let $0 \le \Delta_\omega(v, v^\prime) \le n$ be their differential Hamming weight or Hamming weight difference, i.e., $\Delta_\omega(v, v^\prime) \triangleq \lvert\omega(v) - \omega(v^\prime)\rvert$. $\uparrow\Delta_\omega(v, v^\prime)$ and $\downarrow\Delta_\omega(v, v^\prime)$ represents {\itshape maximum} and {\itshape minimum} differential Hamming weights between $v$, $v^\prime$, respectively.
\end{definition}

\noindent
The notion of {differential Hamming weight} that we have coined in this work is subtly different from that of {Hamming distance}, which is a metric for measuring the edit distance between two sequences, $d(x, y) \triangleq \omega(x \oplus y)$. For instance, $d(\langle 11110000 \rangle_U, \langle 00001111 \rangle_U) = 8$ whereas their $\Delta_\omega$ is 0.

Cryptographic implementations in hardware, such as an AES accelerator in a microcontroller where the algorithm is not running as a software process on a dedicated hardware accelerator, are much more likely to be vulnerable to the {Hamming distance leakage}. Since they typically have only a few interconnections between registers, this leads to a detectable Hamming distance as opposed to a Hamming weight. Therefore, {Hamming distance leakage model} is commonly used in power or EM attacks on hardware implementations~\cite{hw-hack-book} of cryptographic algorithms, while {Hamming weight leakage model} targets software implementations running on a microcontroller or a microprocessor. \ferhat{Nevertheless, \tool additionally employs {Hamming distance leakage model} in its symbolic analysis.}

A standard {\itshape safety} property states that nothing bad can happen along one execution trace; however, information flow properties relate two execution traces and called 2-hypersafety properties in literature~\cite{hyperproperties}.
Differential Hamming weight and Hamming distance will be used in detection of {\itshape leakage points} (cf. \S\ref{sec:diff-sym-analysis})
and can be thought as an interpretation of \emph{2-hypersafety property} in power side-channel analysis. If the Hamming weight of two different values is the same or their difference is close to zero, then it is very hard for an attacker to differentiate them from the power traces and there is likely no vulnerability.
On the other hand, if the Hamming weight of two different values has large difference, then it is much more likely an attacker can detect these different values from the power traces.

Let's assume that $k_1$ and $k_2$ are sensitive 8-bit inputs to a function $c \leftarrow F(k)$ of a cryptographic code 
and that their Hamming weights are $\omega(k_1)=8$ and $\omega(k_2)=0$. In this case, their $\Delta_\omega$ is maximum possible since for 8-bit vectors ($\mathbb{F}_8$) there cannot be bigger difference than 8. If we aligned averaged power traces of two executions, $F(k_1)$ and $F(k_2)$, we would expect to see a distinguishing spike at the point where $F(k_1)$ and $F(k_2)$ are executed, from Definition~\ref{def:monotone}, i.e. $\lvert\overline{P}(F(k_1)) - \overline{P}(F(k_2))\rvert > 0$.
In Figure~\ref{fig:hw-vs-delta-hw} on the right graph, such spikes can be seen. Those power traces are collected from an ARM Cortex-M4 microcontroller's 32-bit target architecture to profile data-oriented differences on $asr$ instructions on the same cryptographic implementation. The most distinguishable Hamming weight differences are those closer to 8 and the least distinguishable ones are those closer to 0.

\revision{
  \begin{definition}[Formal Leakage Definition]
    Let $P$ be a program with a function $f (k, x) = c$, where $k$ is the sensitive (secret) input, $x$ is the public input, and $c$ is the output. We define a leakage in the binary of $f$ if there exist two sensitive input vectors $k_1$ and $k_2$, that slice the secret input domain of $k$ into two subsets at the destination register of location $i$ with a distinguishable Hamming weight difference. Therefore, $P$ has a leakage if $\forall x, {k_1} \in \vec{k_1}, {k_2} \in \vec{k_2}, \lvert \omega(r_i(k_1, x)) - \omega(r_i(k_2, x))\rvert \le \nu$ where $\omega$ is a function that returns Hamming weight of the register $r$ and $\nu$ is a threshold to distinguish hamming weight difference. $\nu$ is automatically set with the maximum width that the register $r$ can take at location $i$. Those points can be easily exploited by the attacker through a single-trace attack.~\rebuttal{\#5}
  \end{definition}
}


\ferhat{
  We conducted a literature review on the recent power/EM side-channel attacks (cf. Table~\ref{tab:attack-classification} in Appendix~\ref{sec:discussion-of-existing-vulnerabilities}). We observe some commonalities among those \emph{single-trace side-channel attacks}~\cite{2021:HOST,2021:1307,2021:790,2021:858,2021:104,2020:1559,2020:992,2020:549,2020:368,2019:1236,2018:1,sim2018key,sim2017key,2016:923} and have made two observations:
  \begin{observation}[Case splits are \emph{points of interest}]
    Attackers identify the most vulnerable instruction locations in the binaries as specific \emph{points of interest (PoIs)}, where the domain of an intermediate value is significantly shrunk into a set of values (\emph{cases}). Thus, they only need to observe the power traces at those PoIs to detect the leakage; in fact, they can create templates for each Hamming weight \emph{classes} where those values belong to and then compare the power traces to the templates to detect the leakage using statistical techniques.
  \end{observation}
}

\ferhat{
  \begin{observation}[{Differential Hamming weights} at \emph{points of interests} reflect distinguisable observations]\label{obs:diff-hw}
    Similarly, intermediate values where differential Hamming weight among them become high are of interest since they are highly distinguishable on a power or EM trace.
  \end{observation}
}
\noindent Existing work introduces the notion of {\itshape determiner} in their single-trace attacks on Lattice-based key encapsulation~\cite{2020:992}.

\begin{definition}[Determiner]\label{def:determiner}
  The determiner is an intermediate value that is defined according to a sensitive bit value, and the difference between the Hamming weights of the elements of the determiner domain is greater than or equal to 2. The cardinality of the determiner domain is 2.
\end{definition}

\subsection{Hamming Weight Classes}\label{sec:entropy}
For deterministic systems, the Shannon entropy can also be used to give a measure of the leakage of the side-channel, corresponding to the observation gain (on the secret) after one round of observation~\cite{phan2017synthesis, m2012measuring}.

If an adversary learns the Hamming weight $\omega$ of an 8 bit-width intermediate value $v$, then this reduces the uncertainty about $v$ as only $\binom{8}{w}$ out of 256 values satisfy the observed Hamming weight.
For example, observing an $\omega$ of 4, it gives an attacker 70 possible secret keys, whereas observing an $\omega$ of 0 or 8 leads to only one possible secret key. Accordingly, the occurrence of Hamming weight {\itshape classes} close to 4 are more likely, but bring less information about the secret key~\cite{2018:476}.

\begin{table}[ht!]
  \caption{$\omega$ classes and probabilities for $\mathbb{F}_8$}
  \label{tab:hw-classes}
  \setlength{\tabcolsep}{0.6em}
  \ifccstemplate\small\else\footnotesize\fi
  \begin{tabularx}{\linewidth}{Cccccccccc}
    \toprule
    $\omega_i$                       & 0    & 1    & 2    & 3    & 4    & 5    & 6    & 7    & 8    \\
    \midrule
    $|\omega_i|$                     & 1    & 8    & 28   & 56   & 70   & 56   & 28   & 8    & 1    \\
    \midrule
    $\ferhat{\mathbb{P}_{\omega_i}}$ & .004 & .031 & .109 & .219 & .273 & .219 & .109 & .031 & .004 \\
    \bottomrule
  \end{tabularx}
\end{table}

\ferhat{
  Since \tool~relies on the program's compiled semantics and data encoding at intermediate values (destination registers), it naturally identifies those interesting points where their \emph{Shannon's entropy}~\cite{Borda2011} becomes significantly low. Therefore, we propose an approximate model to obtain the entropy of destination registers 
  to quantify the leakage over \emph{single-run} (or \emph{single-trace}) in the attack.
}

\begin{definition}[\ferhat{$\omega$-class sampling model}]\label{def:entropy}

  It samples each \emph{Hamming weight class} $\omega_i$ of a destination register $r$, and if we obtain a value, we include the probability $\mathbb{P}_{\omega_i}$ of that class (see Table~\ref{tab:hw-classes}) in entropy calculation.
    {
      \ifccstemplate\else\small\fi
      \begin{equation*}
        \tilde{\eta}(r) = -\sum_{i=0}^{n} {\mathbb{P}_{\omega_i}(r) \cdot \log_2 \mathbb{P}_{\omega_i}(r)}, \,\,\mathrm{where} \, r \in \mathbb{F}_n
      \end{equation*}
    }
  \noindent For instance, a register having only all-ones or all-zeros has $\tilde{\eta}$ of 1.00. The maximum entropy for an intermediate value in $\mathbb{F}_n$ is 2.54.
\end{definition}

\section{Approach}
\label{sec:approach}

\begin{figure*}[t]
  \includegraphics[trim={0.02cm 0.2cm 0.05cm 0.21cm}, clip, width=\linewidth]{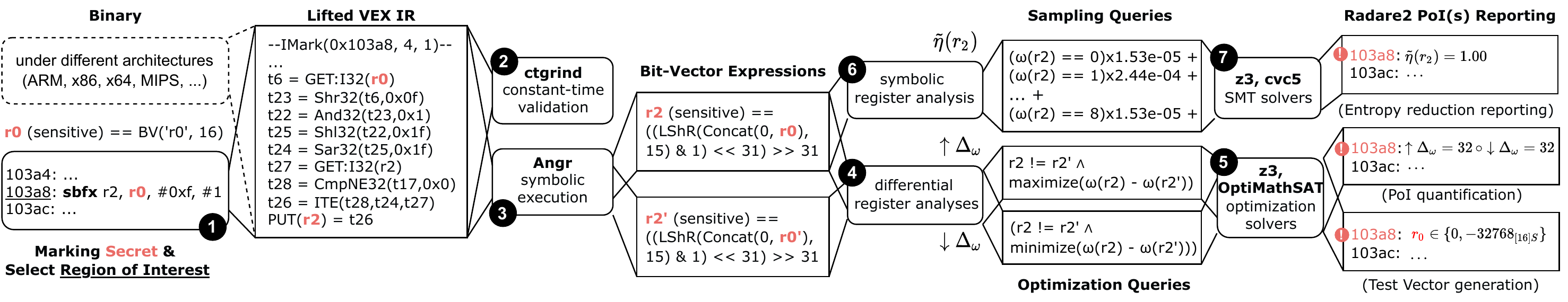}
  \caption{\tool's simplified tooling workflow under \emph{differential Hamming weight} and \emph{$\bm{\omega}$-class sampling} models.}
  \label{fig:tool-workflow}
  \ifccstemplate\Description{}\fi
\end{figure*}

In our formal analysis \ferhat{of binaries}, we will be using the notion of differential Hamming weight, Hamming distance, and Hamming weight classes.

\subsubsection*{Threat Model}\label{sec:threat-model}

Our work mainly focuses on {\itshape single-trace side-channel attacks} against constant-time implementations.
Single-trace side-channel attacks aim to extract a secret value from one side-channel measurement.
First, the attacker has physical access to the identical device and can configure it with known code and secrets to capture power traces. Therefore, the adversary can capture multiple measurements to create {\itshape templates} corresponding to known keys or intermediate values of the target cryptographic algorithm implementation. Second, the adversary has sufficient knowledge about the details of the target software implementation, i.e., the adversary downloads and inspects the publicly available software packages, e.g., candidates submitted to NIST PQC standardization process. When running the attack, the adversary is limited to a single measurement. With single-trace attacks, key generation and encapsulation can be targeted since they use one-time values ({\itshape ephemeral secrets}).

\subsection{Tooling Workflow}\label{sec:implementation}
In this section, we discuss the general workflow of the proof-of-concept implementation of our approach.

A user first needs to specify a function of interest or a starting address in the binary at which secret inputs or registers are defined as symbolic values.
At the stage~\circled{1}~in Figure~\ref{fig:tool-workflow}, the user wants to analyze an ARM binary starting from the \texttt{x103a8} address. This location starts with a \texttt{sbfx} (Signed Bit Field Extract) instruction which copies adjacent bits from one register into the least significant bits of a second register, and sign extends to 32 bits. Register \texttt{r0}, which holds the first operand, is marked as \emph{sensitive} and is defined with a symbolic bit-vector in $\mathbb{F}_{16}$.

Binary analysis tools typically consists of a disassembler and a lifter to disassemble a given binary and translate it into what is known as an Intermediate Representation (IR).
In \tool, we preferred to use \texttt{angr} binary analysis framework (see stage~\circled{3}~in Figure~\ref{fig:tool-workflow}) which operates on the lifted VEX IR, and also provides a convenient API to perform symbolic execution, expression annotation, and static analyses such as \emph{loop finder}.
In Figure~\ref{fig:tool-workflow}, the \texttt{sbfx} instruction (stage~\circled{1}) is lifted into the corresponding set of VEX statements in an \texttt{irsb} block by PyVEX~\cite{pyvex} (stage~\circled{3}).
The VEX is structured into instruction superblocks (\texttt{irsb}). Each \texttt{irsb}
contains a list of statements which may modify the program state and ends with an exit statement describing the succeeding \texttt{irsb} location.
The most common statements are \texttt{WrTmp}, \texttt{Put}, and \texttt{St} to write temporary variables, registers, and memory respectively. Statements use expressions such as \texttt{RdTmp}, \texttt{Get}, and \texttt{Ld} to read temporary variables, registers, and memory~\cite{vexSemantics}. Unary and binary operations such as \texttt{Shr}, \texttt{And}, and \texttt{Xor} have variants based on different bit-widths.
Although, the focus of our research is on analysis of constant-time cryptographic implementations, in our tooling workflow, our method first checks any violations of constant-time practices using \texttt{ctgrind} tool developed by \cite{ctgrind} (stage~\circled{2}) and \tool rejects to continue if any violations are found.

\tool~uses forward symbolic execution~\cite{forwardSymEx} to find min/max Hamming weight differences and \ferhat{approximate entropy of registers} in tandem with satisfiability modulo theory and optimization solvers~\cite{z3, z3opt}, and employs dynamic taint tracking over bit-vector expressions using a dynamic annotation propagation mechanism \ferhat{on bitvector expressions} (provided by \texttt{angr}'s \texttt{claripy} module).
For each instruction that the symbolic execution engine steps over, \tool~identifies secret-tainted, logical and arithmetic instructions.
For each of them, it lazily extracts the destination register's symbolic bit-vector expression from the program's symbolic store (cf. \S\ref{sec:symex}).
\tool~then creates a self-composed version of the expression by introducing a fresh variable for each symbolic variable (cf. \S\ref{sec:diff-sym-analysis}). In this example, \texttt{r2$^\prime$} is introduced after the stage~\circled{3}.
\tool~later creates two optimization queries over the conjunction of those two bit-vector expressions, with the objective of maximizing and minimizing differential Hamming weights or \ferhat{Hamming Distance}, \ferhat{and analyzed them over SMT-based \emph{single-objective, linear optimization} algorithms}~(after stage~\circled{4}).
\tool~records the results of the optimization queries and SMT queries as annotations at the respective address locations in the binary using a popular reverse engineering framework, \texttt{Radare2}~\cite{radare2}. Additionally, if \tool~detects a \emph{determiner} (cf. Definition~\ref{def:determiner}) then
it adds a vulnerability flag to the binary~(after stage~\circled{5}) at the respective location \ferhat{as an indication of \emph{Point of Interest}. It also returns a function to generate a pair of \emph{Test Vectors} to be used as an input for specific \emph{t-test} (TVLA).}
\ferhat{
  For the case of symbolic register analysis (stages~\circled{6} and~\circled{7}), we calculate the approximate entropy using $\omega$-class sampling method (see \S\ref{sec:entropy} and \S\ref{sec:register-analysis}). We developed three SMT-based techniques, parallel, pseudo-boolean equalities, and incremental using \texttt{z3} and \texttt{CVC5} solvers for bit-vector arithmetic.
}

In this example, {\small\texttt{sbfx r2, r0, \#0xf, \#1}} extracts the 16th bit position from register $r_0$, and store it in register $r_2$ by sign extending to 32 bits. Consequently, $r_2$ can only take two values, 0 or -1 ($= 0xFFF\ldots F$). Therefore, the power side-channel analysis at address \texttt{x103a8} results in {\small$\uparrow\Delta_\omega = 32$} and {\small$\downarrow\Delta_\omega = 32$} with \ferhat{$\tilde\eta = 1.00$,} and critical intermediate values for $r_0$ are found as {\small$-32768_{[16]S}$} and 0.

\subsection{Register Analysis Techniques}\label{sec:symex}
\ferhat{
  The principle of~\tool's verification scheme depends on analyzing each register that is tainted by the secret input(s) at every point of execution of the code.
}

\tool~uses \emph{forward symbolic execution} and \emph{dynamic taint analysis} to perform \ferhat{\emph{symbolic register analysis} at binary level}:
{dynamic taint analysis} runs a program and observes which computations are affected by predefined taint sources such as secret function inputs; and
  {forward symbolic execution} automatically builds a logical formula describing registers at each state of the execution, which reduces the problem of reasoning about the execution to the domain of bit-vector logic. The two analyses can be used in conjunction to build formulas representing only the parts of an execution that depend upon tainted values~\cite{forwardSymEx}.

\revision{Another purpose of using these techniques together is to quickly identify potential Point-of-Interest (PoI) candidates within the code, and to determine which functions have these candidates. The dynamic taint analysis is used to narrow down the number of potential PoIs without invoking a solver, by identifying which registers are tainted by the secret input. However, dynamic analysis alone is not sufficient to prove the existence of vulnerabilities, as it is not able to show that there are always only two classes of data whose Hamming weight difference is distinguishable. This is where symbolic execution comes into play, by providing symbolic representations of destination registers, extracted from the symbolic states of the computation, to further verify the existence of vulnerabilities.}~\rebuttal{\#2}

Symbolic execution operates on symbolic values that represent any possible concrete value. Some or all variables (or in the context of binary analysis,
registers or memory locations) are represented by a symbol that stands in for any possible value the variable could take. As the execution proceeds, symbolic execution computes logical formulas over these symbols. These formulas represent the operations performed on the symbols during execution and describe limits for the range of values the symbols can represent~\cite{binary-analysis-book}.

The \texttt{angr}'s symbolic execution engine computes two different kinds of formulas over these symbolic values: a set of symbolic expressions and a path constraint. A \emph{symbolic state} $S_n$ consists of the path constraint and a \emph{symbolic store} $\sigma_n$ that maintains a mapping of all registers, temporaries, and memory locations to symbolic expressions.

In \tool, the user introduces taints by annotating the initial symbolic bit-vector values (registers or function parameters) as \emph{sensitive}; the rest of registers, temporaries, and memory locations are initialized as untainted. In Figure~\ref{fig:symex}, symbolic bit-vector $\beta_{[8]}$ denotes the secret (high) parameter $x_0$, and symbolic bit-vector $\lambda_{[8]}$ denotes the untainted (low) parameter ${sum_0}_{[8]}$. Taint propagation is performed dynamically over symbolic bit-vector expressions by relocating taint annotations from one symbolic expression to another while symbolic execution progresses to construct the symbolic store.
  {
    \begin{figure}
      \centering
      \setlength{\tabcolsep}{0.7em}
      \ifccstemplate\else\small\fi
      \begin{tabular}{@{}cll@{}}
        state & lifted binary code                & symbolic store $\sigma$                                                       \\
        \midrule
        $S_0$ & $f({sum_0}_{[8]}, \, x_{0{[8]}})$ & $\sigma_0 := \{\red{x_0} = \red{\beta}_{[8]} \wedge sum_0 = \lambda_{[8]} \}$ \\
        $S_1$ & $r_0:= x_0- 64$                   & $\sigma_1 := \{\sigma_0 \wedge\red{r_0} = \red{x_0} - 64 \}$                  \\
        $S_2$ & $r_1:= r_0 \gg 7 $                & $\sigma_2 := \{\sigma_1 \wedge\red{r_1} = f_{asr}(\red{r_0}, 7) \}$           \\
        $S_3$ & $r_2:= \, \sim r_1 $              & $\sigma_3 := \{\sigma_2 \wedge\red{r_2} = \sim \red{r_1} \}$                  \\
        $S_4$ & $r_3:= r_2 \land x_0$             & $\sigma_4 := \{\sigma_3 \wedge\red{r_3} = \red{r_2} \land \red{x_0} \}$       \\
        $S_5$ & $r_4:= sum_0 + r_3 $              & $\sigma_5 := \{\sigma_4 \wedge\red{r_4} = sum_0 \land \red{r_3} \}$
      \end{tabular}
      \caption{Symbolic execution of the example code.}
      \label{fig:symex}
      \ifccstemplate\Description{}\fi
    \end{figure}
  }

As an illustration, consider the program in Figure~\ref{fig:symex}, for the sake of clarity its static single assignment form will be discussed in this section as a working example instead of a target specific machine code. Our analysis is being performed on disassembled binaries where intermediate values are represented by \emph{registers} and \emph{temporary} locations based on the target instruction set architecture. `Closed quantifier-free formulas over the theory of bit-vectors' (QF\_BV)~\cite{smtlib} is used to represent symbolic expression and formulas.
In Figure~\ref{fig:symex}, first, \tool~assigns symbolic values (unconstrained bit-vector values) to $\beta$ and $\lambda$ to initiate a \emph{call state} and uses standard forward symbolic execution to update the symbolic store until the first secret-dependent critical instructions is met at state $S_2$. Just after the symbolic store updates destination register $r_1$, \tool~queries the symbolic store for the value of the temporary $r_1$ to fetch the corresponding symbolic bit-vector expression and apply variable elimination and simplification:
{\ifccstemplate\else\small\fi $\sigma_2{[r_1]} \mapsto (\beta_{[8]S} - 64) \gg 7$}.
At this point the bit-vector expression can be discharged to an \texttt{SMT} solver to get a satisfying model for symbolic value $r_1$: {\ifccstemplate\else\small\fi $\exists x_{[8]}. bv_{asr}(bv_{sub}(x_{[8]}, 64), 7)$}. However, \tool~does not create a solver instance, just caches the query and its address location in the binary at this stage.

\subsubsection*{\ferhat{Differential Symbolic Register Analysis}}\label{sec:diff-sym-analysis}

Differential analysis aims to find different behaviors in programs or verify $k$-safety (i.e., properties that concern interactions between $k$ program runs)~\cite{differentialProgramAnalysis}. \tool~needs to reason about pairs of execution traces (2-safety) to identify different power side-channel behavior for different inputs on the same program; therefore, we are required to model two execution traces in the same symbolic execution instance.

In principle, 2-safety properties can be reduced to standard safety properties of a self-composed program~\cite{k-safety}.
Similarly, symbolic execution can be adapted to the case of constant-time code following the self-composition principle. Instead of self-composing the program, we rather self-compose the formula with a fresh version of its symbolic variables plus a precondition stating that the low inputs are equal:
{
\small
\begin{equation*}
  \begin{split}
    \lambda = \lambda^\prime \wedge
    \begin{pmatrix}
      \; x_0 = \beta \wedge sum_0 = \lambda \wedge r_0 = x_0 - 64 \wedge  r_1 = r_0 \gg 7 \; \wedge \\
      x_0^\prime = \beta^\prime \wedge sum_0^\prime = \lambda^\prime \wedge r_0^\prime = x_0^\prime - 64 \wedge  r_1^\prime = r_0^\prime \gg 7
    \end{pmatrix}
  \end{split}
\end{equation*}
}
However, this can be achieved in a lazy manner by querying the symbolic store to get the symbolic expression of $r_1$ and then self-compose it with a version of itself ($r_1^\prime$) in which all symbolic values are fresh. Since we aim to find a data-oriented difference in a pair of executions, we assume $r_1$ and $r_1^\prime$ are different. In this way, the side-channel formula $\varphi_{SC_{S_2}}$ in state $S_1$ for the destination register $r_1$ can be then simplified to the following:
{
\ifccstemplate\else\small\fi
\begin{equation*}
  \varphi_{SC_{S_2}} \triangleq
  {\overbrace {r_1 \ne r_1^\prime}^{\mathclap{\text{disjoint 2-secrets}}}} \wedge
  {\underbrace{r_1 = (\beta - 64) \gg 7}_{\mathclap{\text{symbolic register }r_1}}} \wedge
  {\overbrace {r_1^\prime = (\beta^\prime - 64) \gg 7}^{\mathclap{\text{self-composition of }r_1}}}
\end{equation*}
}

Once the symbolic execution reaches at state $S_4$, a tainted $and$ instruction is also detected and \tool~caches this symbolic expression of $m_3$ and self-compose it as well.

In literature, there is yet no definitive way of maintaining \emph{constant power-consumption} in software implementations similar to that of \emph{constant-time} due to the data dependent power consumption of digital circuits (see \S\ref{sec:power-analysis}).
Therefore, it is not possible to reduce this problem to a \emph{decision problem}. Nevertheless, we can still use {differential program analysis} to model the power side-channel vulnerabilities by introducing an $\omega$ constraint to check if there are two intermediate values $r_1$ and $r_1^\prime$ whose differential Hamming weight is greater than a certain threshold (e.g., {\small$\lvert\omega(r_1) - \omega(r_1^\prime)\rvert > \nu_{asr}$} for \texttt{asr} instruction). However, this query would be only useful for an attacker to craft low inputs to gain an observable difference in power traces, and can be used in \emph{chosen-ciphertext} attacks, but it is not useful to find a definitive weak point such as a \emph{determiner} (see Definition~\ref{def:determiner}) in the binary and does not explain much in terms of quantification of power side-channel vulnerability. Therefore, we introduce minimum and maximum differential Hamming weight (see Definition~\ref{def:diff-hw}) inspired from differential power analysis (see \S\ref{sec:diff-power-analysis}) to pinpoint the large \emph{Hamming weight swings} happened among a constrained set of values.

Since those binary locations where small number of large Hamming weight swings happen are likely to be the locations where attackers can easily gain observable differences in power traces, we introduce a measure to quantify the power side-channel vulnerabilities using differential Hamming weight (see Definition~\ref{def:diff-hw}), differential Hamming distance (see Definition~\ref{def:hd}) and Hamming weight class sampling (see Definition~\ref{def:entropy}). If the maximum $\Delta_\omega$ is equal to minimum $\Delta_\omega$ or their difference ($\lvert \uparrow\Delta_\omega - \downarrow\Delta_\omega\rvert$) is significantly low considering the bitwidth of the intermediate values, then the implementation can be considered as not robust against power side-channel attacks. Because an attacker can easily generate a few templates for respective $\omega$ classes of those intermediate values (e.g. Table~\ref{tab:hw-classes} for $\mathbb{F}_8$).

To use in bit-vector expressions, we introduce a Hamming weight function $\omega$ in the theory of bit-vectors in the \texttt{SMT} solver as $\omega: \mathbb{F}_n \mapsto \mathbb{F}_m$ where $n$ is the width of the input vector and $m = \lfloor \log_2 n \rfloor$ is the width of its Hamming weight. SMT solvers perform better at bit-vector arithmetic in smaller bit-widths due to bit-blasting, therefore, we aim here to have a smaller bit-width for $\omega$ function.
The following shows the objective functions for the optimization problems constructed by \tool~at state $S_2$ (other states are similar).
\begin{equation*}
  \centering
  \setlength{\tabcolsep}{0.6em}
  \ifccstemplate\else\small\fi
  \begin{tabular}{@{}lll@{}}
    state                                      & $\Delta_\omega$ objective function        & optimization query                                                               \\
    \midrule
    \multicolumn{1}{c}{\multirow{2}{*}{$S_2$}} & maximize $\Delta_\omega(r_1, r_1^\prime)$ & $\varphi_{SC_{S_2}} \land max(\lvert\omega({r_1}) - \omega({r_1}^\prime)\rvert)$ \\
                                               & minimize $\Delta_\omega(r_1, r_1^\prime)$ & $\varphi_{SC_{S_2}} \land min(\lvert\omega({r_1}) - \omega({r_1}^\prime)\rvert)$ \\
  \end{tabular}
\end{equation*}
After discharging those queries into the optimization solver, we can see the quantification results in Figure~\ref{fig:analysis-result-sketch}. At state $S_2$, the difference of differential Hamming weight is minimized, and at state $S_4$ difference of differential Hamming weight is maximized. The former is marked as vulnerable, and the latter is not. In addition, a pair of witness is generated for the vulnerable state.

\begin{figure}
  \centering
  \setlength{\tabcolsep}{0.45em}
  \ifccstemplate\else\small\fi
  \begin{tabular}{@{}clccc@{}} 
          & machine code                                 & $\tilde{\eta}$($entropy$)         & $\Delta_\omega$($weight$)                           & $d$($distance$)                           \\
    \midrule
    $S_1$ & $r_0:= x_0 - 64$                             & $\tilde{\eta}(r_0) = 2.54$        & $\Delta_{\uparrow\downarrow\Delta_\omega}=8$        & $\Delta_{\uparrow\downarrow\ d}=7$        \\
    $S_2$ & $r_1:= r_0 \gg 7  \quad\red{\star}$          & $\tilde{\eta}(r_1) = 1.00$        & $\Delta_{\uparrow\downarrow\Delta_\omega}=0$        & $\Delta_{\uparrow\downarrow\ d}=0$        \\
    $S_3$ & $r_2:= \, \sim r_1 \:\quad\quad\gray{\star}$ & $\gray{\tilde{\eta}(r_2) = 1.00}$ & $\gray{\Delta_{\uparrow\downarrow\Delta_\omega}=0}$ & $\gray{\Delta_{\uparrow\downarrow\ d}=0}$ \\
    $S_4$ & $r_3:= r_2 \land x_0$                        & $\tilde{\eta}(r_3) = 2.54$        & $\Delta_{\uparrow\downarrow\Delta_\omega}=8$        & $\Delta_{\uparrow\downarrow\ d}=7$        \\
    $S_5$ & $r_4:= sum_0 + r_3$                          & $\tilde{\eta}(r_4) = 2.54$        & $\Delta_{\uparrow\downarrow\Delta_\omega}=8$        & $\Delta_{\uparrow\downarrow\ d}=7$        \\
  \end{tabular}
  \caption{Three different leakage analysis results ---Point-of-Interests are marked with stars.}\label{fig:analysis-result-sketch}
  \ifccstemplate\Description{}\fi
\end{figure}

After the first vulnerability is identified at state $S_2$, the second vulnerability would have been also identified at state $S_3$. However, \tool~does not trigger another query at this point, instead it uses the $S_2$'s solver instance to evaluate the min/max $\Delta_\omega$ at $S_3$. The reason is the first vulnerability that happens at state $S_2$ affects the state $S_3$ since the symbolic expression that represents $r_2$ is a logical implication of $r_1$. We call this as {\itshape leakage continuity effect}, and it occurs when consecutive instructions continue operating on the reduced domain, which leads to more observable signal in the power trace.

Since optimization queries are more expensive than standard SMT queries and \tool~also aims at providing solutions for critical intermediate values (at least two {\itshape witnesses}) to generate Test Vector at each PoI for further TVLA analysis, we use Algorithm~\ref{algo} to reduce the number of queries once dynamic taint propagation encounters an instruction whose VEX interpretation uses one of the critical instructions in a symbolic state.

In Algorithm~\ref{algo}, $r$ is the bit-vector expression of the register (or temporary) under analysis and $r^\prime$ is its self-composed version. The algorithm first initializes the {\itshape objective} function at Line 1, sets minimization objective at Line 2, and finds the maximum $\Delta_\omega(r, r^\prime)$ at Line 4. Function \texttt{push} and \texttt{pop} are to retain solver instance's internal state in successive queries. \texttt{model} checks the satisfiability of a (optimal) solution, and if it is satisfiable (\texttt{SAT}), it returns a {\itshape finite model}. Function \texttt{eval} evaluates the given expression based on the {\itshape model} found. If the $\downarrow\Delta_\omega$ equals the maximum Hamming weight difference possible considering the bit-width of the expression $m$ (Line 6) then there is no need to find $\uparrow\Delta_\omega$ since they are both equal; otherwise, it discharges maximization objective at Line 10. Function \texttt{flag} marks the address of the instruction under analysis as vulnerable in the binary. At line 13, solver assigns solutions to the two witnesses. We call the constraint constructed at line 16 as {\itshape discriminant}, and it is to enforce values of two witnesses at bit position $n$ to be disjoint (1 or 0). It is used to check if there are other solutions when $\uparrow\Delta_\omega$ equals to $\downarrow\Delta_\omega$ at a Hamming weight class $n$ at line 17. If the result is unsatisfiable (\texttt{UNSAT}) then there are no other solutions in this $\omega$ class, and the instruction address is flagged as vulnerable. The case analysis at line 19 is to capture edge scenarios such as a mask is crafted on purpose to have the same Hamming weight (e.g. $w_1 = 0x5A3C$ and $w_2 = 0xA5C3$) but having maximum {\itshape Hamming distance} to make it difficult for an attacker to change one valid value to a different valid value through {\itshape fault-injection} attacks. It is a defense used by smart card industry~\cite{witteman2008secure}.
However, this case hasn't showed up yet in our analysis of PQC candidates.

\begin{algorithm}[t]
  \footnotesize
  \caption{\small Simplified version of differential symbolic register analysis at stage~\circled{4} in Figure~\ref{fig:tool-workflow}}\label{algo}
  \SetKwData{WitnessFirst}{$w_1$}
  \SetKwData{WitnessSecond}{$w_2$}
  \SetKwData{Min}{$\downarrow\Delta_\omega$}
  \SetKwData{Max}{$\uparrow\Delta_\omega$}
  \SetKwData{None}{$none$}
  \SetKwData{SAT}{SAT}
  \SetKwData{UNSAT}{UNSAT}
  \SetKwData{Empty}{$\emptyset$}
  \SetKwData{Solver}{$solver$}
  \SetKwData{R}{$r$}
  \SetKwData{Rp}{$r^\prime$}
  \SetKwData{Xor}{$\oplus$}
  \SetKwData{Gets}{$\gets$}
  \SetKwData{Objective}{$objective$}
  \SetKwData{Vulnerable}{vulnerable}
  \SetKwData{Instruction}{$instruction$}
  \SetKwFunction{Model}{solve}
  \SetKwFunction{Push}{solver.push}
  \SetKwFunction{Pop}{solver.pop()}
  \SetKwFunction{Minimize}{solver.minimize}
  \SetKwFunction{Maximize}{solver.maximize}
  \SetKwFunction{Eval}{solver.eval}
  \SetKwFunction{Add}{solver.add}
  \SetKwFunction{Flag}{flag}
  \SetKwFunction{Width}{width}
  \SetKwInOut{Ensure}{Ensure}
  \Ensure{for each secret-tainted VEX instruction}
  \KwData{\R, \Rp : bit-vector expressions $\in \mathbb{F}_n$; \Solver: solver \emph{context}}
  \KwResult{($\uparrow\Delta_\omega$, $\downarrow\Delta_\omega$); (\WitnessFirst: witness\textsubscript{1}, \WitnessSecond: witness\textsubscript{2})}
  (\Max, \Min) \Gets (\None, \None)\; 
  \Objective \Gets $\mid\omega(\R) - \omega(\Rp)\mid$\;
  \Push{\Minimize{\Objective}}\;
  \lIf{\Model{\Solver} $=$ \SAT}{\Min \Gets \Eval{\Objective}}
  \nl\Pop\;
  \eIf(\tcc*[f]{when \Min equals to width of $r$}){\Min $= n$}{
    \Max \Gets \Min; \Flag{\Vulnerable}\;
  }{
    (\WitnessFirst, \WitnessSecond) \Gets (\None, \None)\;
    \Push{\Maximize{\Objective}}\;
    \If{\Model{\Solver} $=$ \SAT}{
      \Max \Gets \Eval(\Objective)\;
      (\WitnessFirst, \WitnessSecond) \Gets \Eval{\R, \Rp}\;
    }
    \Pop \;
    \If{\Min $=$ $i$ $\land$ \Max $=$ $i$, \emph{where $i$ is an $\omega$-class}}{
      \Push{\Add{$\R[n]$ \Xor $\Rp[n]$}}\;
      \lIf{\Model{\Solver} $=$ \UNSAT}{\Flag{\Vulnerable}}
      \Pop\;
      \begin{minipage}[t]{1\linewidth}
        \If{\WitnessFirst $\neq$ \None $\land$ \WitnessSecond $\neq$ \None $\land$ $\omega(\WitnessFirst) = \omega(\WitnessSecond)$}{
          \nl \Push{\Add{\R $\neq$ \WitnessFirst $\land$ \Rp $\neq$ \WitnessSecond}}\;
          \nl  \lIf{\Model{\Solver} $=$ \UNSAT}{\Flag{\Vulnerable}}
          \nl  \Pop\;
        }
      \end{minipage}
    }
  }
\end{algorithm}

\subsubsection*{Entropy-based Symbolic Register Analysis}\label{sec:register-analysis}

\ferhat{\emph{Hamming weight-class sampling model} calculates the approximate entropy at destination registers for each secret-tainted arithmetic and logical instruction under analysis. It is a good indicator of {Point of Interests} in \emph{single-trace side-channel attacks}. However, it is computationally expensive and therefore is used as an alternative model.
  In Algorithm~\ref{algo-1}, $r$ is the symbolic bitvector expression, for instance, for $r_1$ in Figure~\ref{fig:symex} it is {\ifccstemplate\else\small\fi $\sigma_2{[r_1]} \mapsto (\beta_{[8]S} - 64) \gg 7$}. Based on Definition~\ref{def:entropy},
  \texttt{normalize} normalizes the discrete distribution of given probability values; additionally, if there is only one value in the \texttt{distribution}, the \texttt{normalize} function takes two more samples in that class to check if the entropy should be normalized to 1.00 or not. Finally, function \texttt{entropy} calculates the entropy of a distribution for given probability values (\texttt{distribution}). We added \emph{blocking constraints} at Line 7 to increase the performance of the solver.}

\begin{algorithm}[t]
  \footnotesize
  \caption{\small Simplified version of symbolic register analysis for \emph{$\omega$-class sampling model} at stage~\circled{6} in Figure~\ref{fig:tool-workflow}}\label{algo-1}
  \SetKwData{WitnessFirst}{$w_1$}
  \SetKwData{WitnessSecond}{$w_2$}
  \SetKwData{None}{$none$}
  \SetKwData{SAT}{SAT}
  \SetKwData{UNSAT}{UNSAT}
  \SetKwData{Empty}{$\emptyset$}
  \SetKwData{Solver}{$solver$}
  \SetKwData{R}{$r$}
  \SetKwData{Xor}{$\oplus$}
  \SetKwData{Gets}{$\gets$}
  \SetKwData{Objective}{$objective$}
  \SetKwData{Distribution}{distribution}
  \SetKwFunction{AddToDistribution}{\Distribution.add}
  \SetKwData{Vulnerable}{vulnerable}
  \SetKwData{Instruction}{$instruction$}
  \SetKwFunction{Model}{solve}
  \SetKwFunction{Push}{solver.push}
  \SetKwFunction{Pop}{solver.pop()}
  \SetKwFunction{Minimize}{solver.minimize}
  \SetKwFunction{Maximize}{solver.maximize}
  \SetKwFunction{Eval}{solver.eval}
  \SetKwFunction{Add}{solver.add}
  \SetKwFunction{Flag}{flag}
  \SetKwFunction{Eta}{$\tilde{\eta}$}
  \SetKwFunction{Width}{width}
  \SetKwFunction{Normalize}{normalize}
  \SetKwFunction{Entropy}{entropy}
  \SetKwInOut{Ensure}{Ensure}
  \Ensure{for each secret-tainted VEX instruction}
  \KwData{\R: bit-vector expression $\in \mathbb{F}_n$; \Solver: solver \emph{context}}
  \KwResult{\Eta of \R}
  \Eta \Gets \None; \Distribution \Gets $\emptyset$\;
  \For(\tcc*[f]{for each $\omega_i$ in $\omega$-classes}){i \Gets 0 \emph{\KwTo} $n$}{
    \Push{\Add{$\omega(\R)=i$}}\;
    \If(\tcc*[f]{sample $\omega$-class}){\Model{\Solver} $=$ \SAT}{\AddToDistribution{$\mathbb{P}_{\omega_i}$};}
    \Pop \;
    \Add{$\omega(\R) \neq i$} \tcc*[r]{block formula}
  }
  \Normalize{\Distribution} \tcc*[r]{may resample at most twice}
  \Eta \Gets \Entropy{\Distribution} \tcc*[r]{see Definition~\ref{def:entropy}}
  \lIf{$\Eta \le 1.00$}{\Flag{\Vulnerable}}
\end{algorithm}


\subsection{Limitations}\label{sec:limitations}

\ferhat{
  We currently individually analyze functions or regions. Pascal's API allows for performing an analysis between given start and end addresses in a given binary file thus any code between start and end address can be analyzed, this will typically be the body of a function, including loops, branches, etc. However, due to the computational complexity of solving symbolic optimization queries for large operations, we generally divide the analysis into sub parts and analyze each sub part (usually functions).

  \revision{
    \texttt{Angr} framework provides the capability to replace the actual code with function summaries. This feature was primarily used to skip analysis on certain portions of the code where we can be certain that there are no leakages. However, it is important to note that our approach does not solely rely on function summaries and they were only used sparingly in our empirical evaluation. A well-known drawback of using function summaries is that it requires careful verification of the correctness of the written summaries, as it may result in false negatives if not properly checked. 
    ~\rebuttal{\#6}
  } 
}

\tool~ performs bounded analysis of loops. Specific loops can be chosen, and loop count can be set by the user through \texttt{angr}'s \texttt{LoopSeer} symbolic execution exploration technique.

\section{Evaluation}\label{sec:analysis-results}

Our evaluation is focused on answering the following research questions:
\begin{enumerate*}[label={{\bfseries{RQ\arabic*}}:}]
  \item Is the proposed approach able to identify {existing vulnerabilities}?
  \item Is the proposed approach able to detect {new vulnerabilities}?
  \item What is the performance of \tool's register analysis methods under different algorithms?
  \item What is the precision of \tool in analyzing protected implementations with randomization and shuffling?
  \item \revision{Show \tool works in a variety of microarchitectures.}~\rebuttal{\#1.b}
\end{enumerate*}

\subsubsection*{Detection of Known Vulnerabilities}\label{sec:vulnerabilities}
We performed a literature review of recent power side-channel attacks against publicly available constant-time implementations of post-quantum cryptographic algorithms submitted to NIST's post-quantum cryptography standardization process, mbedTLS, and the NIST lightweight crypto competition.
In the literature we found $16$ known power-side channel vulnerabilities in constant-time implementations in a variety of cryptographic schemes, from Elliptic Curve~\cite{ecc-based}
and Lattice-based~\cite{lattice-based} to Lightweight~\cite{2021:nist} cryptography. Our method successfully identifies all of them.
We present the results in Table~\ref{tab:attack-classification} \ferhat{in Appendix~\ref{sec:discussion-of-existing-vulnerabilities} and include some of them in our benchmarks (see Table~\ref{tab:evaluation})}.

In our review, we exclude some attack papers that fall under the following categories:
\begin{enumerate*}[nosep, leftmargin=*, label={(\arabic*)}]
  \item classic attacks to block ciphers such as AES's \texttt{sbox}~\cite{2021:611, 2020:TCHES}. One of the reason is that employing {table look-ups} indexed by secret data is not a recommended constant-time practice and in those attacks although Hamming weight or Hamming distance leakage models are being used, the attacker needs to profile all the Hamming weight classes of intermediate values from Galois Field, $\mathbb{GF}(2^8)$, to the \texttt{sbox} function;
  \item timing side-channel attacks using power traces such as~\cite{2017:505, 2017:594} since there are many formal constant-time analysis tools that is able to detect secret-dependent branching;
  \item primitives protected with perfect masking such as order-d secret-sharing masking scheme~\cite{prouff2013masking} considering the fact that the vulnerability is already eliminated by the masking technique.
\end{enumerate*}

\newcounter{program}
\newcommand{\inc}{\stepcounter{program}$\text{P}_\text{\theprogram}$}

\newcommand{\cWork}{\Gape[-2pt]{\thead{\bfseries{Attack}}}}
\newcommand{\cType}{\Gape[-2pt]{\thead[l]{\bfseries{Crypto}\\\bfseries{Type}}}}
\newcommand{\cModel}{\Gape[-2pt]{\thead{\bfseries{Leakage}\\\bfseries{Model(s)}}}}
\newcommand{\cDesc}{\Gape[-2pt]{\thead[l]{\bfseries{Countermeasure}\\\bfseries{Description}}}}
\newcommand{\cOper}{\Gape[-2pt]{\thead[l]{\bfseries{Operation}\\\bfseries{Description}}}}
\newcommand{\cAlgo}{\Gape[-2pt]{\thead[l]{\bfseries{Crypto}\\\bfseries{Algorithm}}}}
\newcommand{\cFunc}{\Gape[-2pt]{\thead[l]{\bfseries{Crypto}\\\bfseries{Function}}}}
\newcommand{\cLoC} {\Gape[-2pt]{\thead{\bfseries{\# of}\\\bfseries{LoC}}}}
\newcommand{\cInst}{\Gape[-2pt]{\thead{\bfseries{\# of}\\\bfseries{Inst.}}}}
\newcommand{\cTime}{\Gape[-2pt]{\thead{\bfseries{Analysis}\\\bfseries{Time (s)}}}}

\begin{table*}[t]
  \caption{Known and New Vulnerabilities. All known attacks in literature levereges {Hamming weight leakage model}.}
  \label{tab:evaluation}
  \centering
  \ifccstemplate\small\else\footnotesize\fi
  \begin{threeparttable}
    \begin{tabular}{@{}|l||l|l|l|l|l||c|c|@{}}
      \hline
           & \cAlgo   & \cType         & \cFunc                            & \cOper                                             & \bfseries{Implementation} & \cWork                     & \cModel       \\
      \hline
      \hline
      \inc & NTRU     & Lattice-based  & \texttt{mod3}                     & Modular reduction                                  & NIST PQC Round3           & \cite{2021:790}            & $\omega$      \\\hline
      \inc & NTRU     & Lattice-based  & \lstinline|mod3_alt|              & Alternative modular reduction                      & NIST PQC Round3           & \cite{2021:790}            & $\omega$      \\\hline
      \inc & NTRU     & Lattice-based  & \lstinline|poly_Z3_to_Zq|         & Map from $\mathbb{Z}_3$ to $\mathbb{Z}_q$          & NIST PQC Round3           & \cite{2020:992}            & $\omega$      \\\hline
      \inc & NTRU     & Lattice-based  & \lstinline|int32_MINMAX|          & Sorting/Comparison                                 & NIST PQC Round3           & \cite{2021:HOST}           & $\omega$      \\\hline
      \inc & NTRU     & Lattice-based  & \lstinline|int32_MINMAX|$^\prime$ & Sorting/Comparison (inline assembly)               & pqm4~\cite{pqm4} library  & ---                        & $\omega$      \\\hline
      \inc & Kyber    & Lattice-based  & \lstinline|poly_frommsg|          & Message Encoding (NIST standard)                   & NIST PQC Round3           & \cite{2021:1307, 2020:549} & $\omega$      \\\hline
      \hline
      \inc & Kyber    & Lattice-based  & \lstinline|poly_frommsg|$^1$      & Message encoding with multiplication               & NIST PQC Round3           & \cite{2020:368}            & $\omega$      \\\hline
      \inc & Kyber    & Lattice-based  & \lstinline|poly_frommsg|$^2$      & Data independent poly. generation                  & NIST PQC Round3           & \cite{2021:1307}           & $\omega$      \\\hline
      \inc & Kyber    & Lattice-based  & \lstinline|poly_frommsg|$^3$      & Balanced data independent poly. gen.               & NIST PQC Round3           & \cite{2021:1307}           & $\omega$      \\\hline
      \inc & Kyber    & Lattice-based  & \lstinline|poly_frommsg|$^4$      & Polynomial randomization                           & NIST PQC Round3           & \cite{2021:1307}           & $\omega$      \\\hline
      \inc & Kyber    & Lattice-based  & \lstinline|poly_frommsg|$^5$      & Byte and bit level random ordering                 & NIST PQC Round3           & \cite{2021:1307}           & $\omega$      \\\hline
      \hline
      \inc & Kyber    & Lattice-based  & \lstinline|poly_frommsg|$^\prime$ & Alternative Message Encoding                       & pqm4~\cite{pqm4} library  & \cite{2020:992, 2020:912}  & $\omega$      \\\hline
      \inc & Kyber    & Lattice-based  & \lstinline|poly_tomsg|            & Convert polynomial to 32-byte message              & NIST PQC Round2           & \cite{2020:1559}           & $\omega$      \\\hline
      \inc & Kyber    & Lattice-based  & \lstinline|mont_reduce|           & Montgomery reduction                               & NIST PQC Round3           & ---                        & $\omega$, $d$ \\\hline
      \inc & Kyber    & Lattice-based  & \lstinline|poly_csubq|            & \lstinline|csubq| of each coefficient a polynomial & NIST PQC Round3           & New                        & $\omega$      \\\hline
      \inc & NewHope  & Lattice-based  & \lstinline|poly_tomsg|            & Convert polynomial to 32-byte message              & NIST PQC Round2           & \cite{2020:1559}           & $\omega$      \\\hline
      \inc & NewHope  & Lattice-based  & \lstinline|poly_tobytes|          & Serialization of a polynomial                      & NIST PQC Round2           & New                        & $\omega$, $d$ \\\hline
      \inc & NewHope  & Lattice-based  & \lstinline|poly_from_msg|         & Convert 32-byte message to polynomial              & NIST PQC Round1           & \cite{2020:368, 2020:549}  & $\omega$      \\\hline
      \inc & FrodoKEM & Lattice-based  & \lstinline|key_decode|            & Message Decoding                                   & NIST PQC Round2           & \cite{2020:1559}           & $\omega$      \\\hline
      \inc & FrodoKEM & Lattice-based  & \lstinline|key_encode|            & Message Encoding                                   & NIST PQC Round3           & \cite{2020:992}            & $\omega$      \\\hline
      \inc & mbedTLS  & Elliptic Curve & \lstinline|ct_mpi_uint_lt|        & Constant-time less-than comparison                 & mbedTLS v3.1.0            & New                        & $\omega$, $d$ \\\hline
      \inc & mbedTLS  & Elliptic Curve & \lstinline|mpi_lt_mpi_ct|         & Constant-time, signed comparison                   & mbedTLS v3.1.0            & New                        & $\omega$, $d$ \\\hline
      \inc & Sparx    & Lightweight    & \lstinline|sparx_encrypt|         & Sparx's ARX-box Assembly                           & NSA Reference             & \cite{2019:335}            & $\omega$      \\\hline
      \inc & Sparkle  & Lightweight    & \lstinline|ARX|                   & Sparkle's ARX-box Assembly                         & NIST LWC Finalist         & ---                        & $\omega$, $d$ \\\hline
    \end{tabular}
    \begin{tablenotes}
      \ifccstemplate\small\else\footnotesize\fi
      \item
      \setcounter{program}{6} \inc, \inc, \inc, \inc, and \inc~are protected implementations of CRYSTALS-Kyber's vulnureable message encoding function ($\B{6}$).
    \end{tablenotes}
  \end{threeparttable}
\end{table*}

The experiments are conducted on ARM Cortex-M4 since popular PQC projects such as pqm4~\cite{pqm4} chooses this target.
\tool~was able to detect all known vulnerabilities listed in Table~\ref{tab:evaluation}.
All vulnerabilities are experimentally shown to be exploitable in the literature and are successfully detected and quantified by~\tool.
Here, we discuss three power side-channel attacks against a lightweight crypto~\cite{2019:335} and a post-quantum crypto~\cite{2020:992,2021:1307} implementations.
More examples have been shown in Appendix~\ref{sec:discussion-of-existing-vulnerabilities}.


SPECK~\cite{2013:404} and SPARX~\cite{sparx} are lightweight block ciphers that use add-rotate-xor (ARX) constructions.  Other examples for ARX constructions include stream cipher Chacha20~\cite{chacha20} and the SKEIN~\cite{skein} hash function.
\cite{2019:335} examines the intuition that ARX ciphers~\cite{sparx} have intrinsic resilience against side channel attacks because of the absence of strong S-Boxes. They performed a correlation power analysis attack to recover the secret. Consider the function \texttt{A} in Listing~\ref{lst:arxbox}.
When compiled with ARM toolchain with \texttt{-O3} and \texttt{cortex-m4} options, it is translated into the assembly code shown in Listing~\ref{lst:arxbox-arm}.
\begin{lstlisting}[language=c, caption={SPECK's ARX-box Implementation}, label={lst:arxbox}, firstline=4, lastline=12, numbers=left, linebackgroundcolor={\ifnum\value{lstnumber}=7\color{light-gray}\fi}]
// Rotate left for 16 bit registers.
#define ROTL(x, n) (((x) << n) | ((x) >> (16-(n))))
// Rotation and Addition.
void A(uint16_t* l, uint16_t* r) {
  (*l) = ROTL((*l), 9);
  (*l) += (*r);
  (*r) = ROTL((*r), 2);
  (*r) ^= (*l);
}
\end{lstlisting}
\tool~detects maximum $\Delta_\omega = 2$ and minimum $\Delta_\omega = 0$. Over the set of $2^{16}$ number, the register \texttt{r3} at line 13 can take only one of $\{0, 1, 2, 3$\} and therefore this reduces the number of traces required for the attack.
\begin{lstlisting}[language={[arm]Assembler}, caption={Full disassembly of Listing~\ref{lst:arxbox}}, numbers=left, label={lst:arxbox-arm}, firstline=4, linebackgroundcolor={\ifnum\value{lstnumber}=10\color{light-gray}\fi}]
ldrh r2, [r0]           ; arg1; 
lsrs r3, r2, 7          ; $\uparrow\Delta_\omega = 9\phantom{0}\;\circ\;\downarrow\Delta_\omega = 0$
orr.w r3, r3, r2, lsl 9 ; $\uparrow\Delta_\omega = 25\;\circ\;\downarrow\Delta_\omega = 0$
uxth r3, r3             ; $\uparrow\Delta_\omega = 16\;\circ\;\downarrow\Delta_\omega = 0$
strh r3, [r0]           ; arg1
ldrh r2, [r1]           ; arg2
add r3, r2
strh r3, [r0]           ; arg1
ldrh r2, [r1]           ; arg2
lsrs r3, r2, 0xe        ; $\uparrow\red{\Delta_\omega = 2}\phantom{0}\;\circ\;\downarrow\red{\Delta_\omega = 0}$
orr.w r3, r3, r2, lsl 2 ; $\uparrow\Delta_\omega = 18\;\circ\;\downarrow\Delta_\omega = 0$
uxth r3, r3             ; $\uparrow\Delta_\omega = 18\;\circ\;\downarrow\Delta_\omega = 0$
strh r3, [r1]           ; arg2
ldrh r2, [r0]           ; arg1
eors r3, r2             ; $\uparrow\Delta_\omega = 18\;\circ\;\downarrow\Delta_\omega = 0$
strh r3, [r1]           ; arg2
bx lr
\end{lstlisting}
NewHope~\cite{newhope} is a key encapsulation method proposed in the NIST post-quantum project. In Listing~\ref{lst:new-hope}, a straightforward implementation might use a for-loop over all message bits containing an if-condition which sets the polynomial coefficients to either 0 or $q/2$. Since such an implementation would be susceptible to timing attacks, the message encoding is implemented in a way that the code inside the for-loop always runs in constant time.

\begin{lstlisting}[language=c, caption={Attack to NewHope's message encoding}, label={lst:new-hope}, firstline=20, lastline=27, numbers=left]
void poly_frommsg(poly *r, unsigned char *msg){
 unsigned int i, j, mask;
 for (i = 0; i < 32; i++) {
   for (j = 0; j < 8; j++) {
     mask = -((msg[i] >> j) & 1);
     r->coeffs[8*i+j+0] = mask & (NEWHOPE_Q/2);
     r->coeffs[8*i+j+256] = mask & (NEWHOPE_Q/2);
     $\cdots$
\end{lstlisting} 
A mask, containing 0 or -1 ($= 0xFFFF\dots$), replaces the if-condition. The mask calculation is shown in Listing~\ref{lst:new-hope} at line 5. However, power consumption might differ between processing a logical zero or logical one, especially because the mask either contains ones or zeroes only. Chances that processed values can be detected by analyzing the power consumption of the device are high.

Based on power measurement,~\cite{2020:368} extracts the complete shared secret from one single trace only. The impact of different compiler directives are additionally investigated: when the code is compiled with optimization turned off (\texttt{-O0} shown in Listing~\ref{lst:new-hope-O0}), the shared secret can be read from an oscilloscope display directly with the naked eye; when optimizations are enabled (\texttt{-O3} shown in Listing~\ref{lst:new-hope-O3-m4}), the attack requires template-based attack, but the attack still works on single power traces.

\begin{lstlisting}[language={[arm]Assembler}, caption={Partial disassembly at \texttt{-O0} of Listing~\ref{lst:new-hope}}, numbers=left, firstline=2, lastline=10, label={lst:new-hope-O0},linebackgroundcolor={\ifnum\value{lstnumber}=7\color{light-gray}\fi}]
ldr   r2, [r7,#0]  ;r2 = memory[r7]
ldr   r3, [r7,#20] ;r3 = memory[r7+20]
add   r3, r2       ;$\uparrow\red{\Delta_\omega} = 32\;\circ\downarrow\red{\Delta_\omega} = 0\phantom{0}\;\Vert\;\red{\tilde{\eta}} = 3.53$
ldrb  r3, [r3,#0]  ;r3 = memory[r3]
mov   r2, r3       ;r2 = r3
ldr   r3, [r7,#16] ;r3 = memory[r3+16]
asr.w r3, r2       ;$\uparrow\red{\Delta_\omega} = 32\;\circ\downarrow\red{\Delta_\omega} = 32\;\Vert\;\red{\tilde{\eta}} = 1.00\quad\red{\star}$
and.w r3, r3, #1   ;$\uparrow\red{\Delta_\omega} = 1\phantom{0}\;\circ\downarrow\red{\Delta_\omega} = 1\phantom{0}\;\Vert\;\red{\tilde{\eta}} = 1.00\quad\gray{\star}$
negs  r3, r3       ;$\uparrow\red{\Delta_\omega} = 1\phantom{0}\;\circ\downarrow\red{\Delta_\omega} = 1\phantom{0}\;\Vert\;\red{\tilde{\eta}} = 1.00\quad\gray{\star}$
\end{lstlisting}

In Listing~\ref{lst:new-hope-O3-m4} at line 2, Arm's {\itshape signed bit field extract} instruction (\texttt{sbfx}) is generated by the compiler for extraction of zero bit (1 bit) of $r2$ and sign-extend it to 32 bits ($\text{if} \; bit_0(r2) == 0, \text{then} \; r2 = 0x0000\cdots, \text{else} \; r2 = 0xFFFF\cdots)$). This instruction's semantics is also similar to right shifting, and actually the VEX IR lifter uses an \texttt{Asr32} instruction in the intermediate representation.

\begin{lstlisting}[language={[arm]Assembler}, caption={Partial disassembly at \texttt{-O3} of Listing~\ref{lst:new-hope}}, numbers = left, label={lst:new-hope-O3-m4}, firstline=1, lastline=7, linebackgroundcolor={\ifnum\value{lstnumber}=2\color{light-gray}\fi}]
ldrb   r2, [r3,#0]    ;r2 = memory[r3]
sbfx   r2, r2, #0, #1 ;$\uparrow\red{\Delta_\omega} = 32\;\circ\downarrow\red{\Delta_\omega} = 32\;\Vert\;\red{\tilde{\eta}} = 1.00\;\;\red{\star}$
strh   r2, r2, #6144  ;r2 = r2 & 6144
and.w  r2, r2, [r0,#0];$\uparrow\red{\Delta_\omega} = 32\;\circ\downarrow\red{\Delta_\omega} = 0\phantom{0}\;\Vert\;\red{\tilde{\eta}} = 3.53$
strh.w r2, [r0,#512]  ;memory[r0 + 512]  = r2
strh.w r2, [r0,#1024] ;memory[r0 + 1024] = r2
strh.w r2, [r0,#1536] ;memory[r0 + 1536] = r2
\end{lstlisting}



\newcommand{\hwmodel}{\multicolumn{4}{c||}{\bf Hamming weight ($\bm{\uparrow\downarrow \Delta_\omega}$)$^\ddag$}}
\newcommand{\hdmodel}{\multicolumn{4}{c||}{\bf Hamming distance ($\bm{\uparrow\downarrow d}$)$^\ddag$}}
\newcommand{\etamodel}{\multicolumn{4}{c||}{\bf Approximate Entropy ($\bm{\tilde\eta}$)$^\ast$}}
\newcommand{\tvla}{\multirow{2}{*}{\bf Pascal}}
\newcommand{\TO}{$^{\red{\mathrm{TO}}}$}
\newcommand{\TP}{\xspace${\green{\mathrm{TP}}}$}
\newcommand{\TN}{\xspace${\green{\mathrm{TN}}}$}
\newcommand{\FP}{\xspace${\red{\mathrm{FP}}}$}
\newcommand{\FN}{\xspace${\red{\mathrm{FN}}}$}
\setcounter{program}{0}
\begin{table*}[t]
  \caption{\tool's Performance Analysis on known and some new vulnerabilities. }
  \label{tab:vulnurebilities}
  \centering
  \ifccstemplate\small\else\footnotesize\fi
  \begin{threeparttable}
    \begin{tabular}{@{}|l||c||c|c|c|c||c|c|c|c||c|c|c|c||c|@{}}
      \cline{2-15}
      \multicolumn{1}{c|}{}        & \bf\# of                        & \hwmodel                        & \hdmodel                        & \etamodel                       & \tvla                                                                                                                                                                                                                                                                              \\
      \hhline{~|~||*{4}{-}||*{4}{-}||*{4}{-}||~|}
      \multicolumn{1}{c|}{}        & \bf \phantom{~}Inst.\phantom{~} & \bf \phantom{~}basic\phantom{~} & \bf \phantom{~}symba\phantom{~} & \bf \phantom{~}obvbs\phantom{~} & \bf \# PoI            & \bf \phantom{~}basic\phantom{~} & \bf \phantom{~}symba\phantom{~} & \bf \phantom{~}obvbs\phantom{~} & \bf \# PoI            & \bf \phantom{~}par.\phantom{~} & \bf \phantom{~}pb.\phantom{~} & \bf $\phantom{~}$inc.$\phantom{~}$ & \bf \# PoI &           \\
      \hhline{-*{14}{=}}      \inc & 77                              & 2.13                            & 2.15                            & 4.49                            & 4                     & 2.01                            & 1.98                            & 4.58                            & 4                     & 5.13                           & 4.86                          & 9.55                               & 30         & \cmark\TP \\\hline
      \inc                         & 73                              & 2.03                            & 2.02                            & 3.89                            & 4                     & 1.89                            & 1.87                            & 4.11                            & 4                     & 4.34                           & 4.14                          & 8.67                               & 31         & \cmark\TP \\\hline
      \inc                         & 52                              & 1.22                            & 1.19                            & 1.60                            & 0                     & 0.64                            & 0.66                            & 1.96                            & 0                     & 3.01                           & 1.98                          & 1.78                               & 0          & \cmark\FN \\\hline
      \inc                         & 46                              & 2.00                            & 1.96                            & 2.00                            & 1                     & 1.40                            & 1.39                            & 2.06                            & 1                     & 5.35                           & 4.41                          & 2.21                               & 1          & \cmark\TP \\\hline
      \inc                         & 24                              & 2.17                            & 2.21                            & 2.21                            & 1                     & 2.18                            & 2.19                            & 2.20                            & 1                     & 2.22                           & 2.25                          & 2.31                               & 1          & \cmark\TP \\\hline
      \inc                         & 64                              & 0.46                            & 0.48                            & 1.03                            & 11                    & 0.40                            & 0.42                            & 1.11                            & 11                    & 1.31                           & 1.32                          & 1.03                               & 11         & \cmark\TP \\\hline
      \hline
      \inc                         & 60                              & 0.51                            & 0.51                            & 0.91                            & 12                    & 0.44                            & 0.45                            & 1.23                            & 12                    & 1.43                           & 1.45                          & 1.19                               & 12         & \cmark\TP \\\hline
      \inc                         & 75                              & 14.07                           & 13.83                           & 15.95                           & 3                     & 13.81                           & 13.77                           & 15.28                           & 3                     & 16.26                          & 16.08                         & 15.00                              & 3          & \cmark\TP \\\hline
      \inc                         & 104                             & 23.88                           & 23.61                           & 27.45                           & 4                     & 23.17                           & 23.27                           & 26.80                           & 4                     & 27.09                          & 26.71                         & 25.32                              & 4          & \cmark\TP \\\hline
      \inc                         & 160                             & 68.61                           & 70.54                           & 81.49                           & 4                     & 75.38                           & 65.37                           & 78.29                           & 4                     & 90.57                          & 91.89                         & 84.24                              & 9          & \cmark\TP \\\hline
      \inc                         & 184                             & 129.03                          & 127.12                          & 146.66                          & 7                     & 123.56                          & 119.94                          & 150.91                          & 7                     & 159.51                         & 153.11                        & 151.64                             & 10         & ---\FP    \\\hline
      \hline
      \inc                         & 59                              & 2.51                            & 2.50                            & 7.27                            & 8                     & 2.06                            & 2.11                            & 6.67                            & 8                     & 7.35                           & 7.51                          & 6.05                               & 8          & \cmark\TP \\\hline
      \inc                         & 139                             & 5.40                            & 5.25                            & 10.81                           & 8                     & 3.60                            & 3.58                            & 9.88                            & 8                     & 21.81                          & 11.49                         & 9.58                               & 8          & \cmark\TP \\\hline
      \inc                         & 48                              & 5.35                            & 6.09                            & 4.82                            & 0                     & 72.59\TO                        & 71.79\TO                        & 69.96\TO                        & 0                     & 48.13                          & 5.81                          & 7.45                               & 0          & ---\TN    \\\hline
      \inc                         & 63                              & 3.83                            & 3.75                            & 8.04                            & 8                     & 2.61                            & 2.53                            & 7.63                            & 8                     & 6.38                           & 8.31                          & 7.04                               & 8          & \cmark\TP \\\hline
      \inc                         & 149                             & 369.77                          & 382.49                          & {139.70}                        & 8                     & 372.65                          & 391.06                          & {127.10}                        & 8                     & 600\TO                         & 600\TO                        & {314.5}                            & 15         & \cmark\TP \\\hline
      \inc                         & 223                             & 5.16                            & 5.56                            & 11.03                           & 8                     & 3.78                            & 3.64                            & 10.17                           & 8                     & 21.48                          & 11.26                         & 10.47                              & 8          & \cmark\TP \\\hline
      \inc                         & 67                              & 24.77                           & 24.48                           & 48.76                           & 16                    & 22.72                           & 22.77                           & 43.89                           & 16                    & 54.52                          & 59.14                         & 45.44                              & 16         & \cmark\TP \\\hline
      \inc                         & 110                             & 161.23                          & 158.77                          & 165.96                          & 3                     & 163.04                          & 160.45                          & 172.02                          & 3                     & 177.01                         & 177.32                        & 170.36                             & 8          & \cmark\TP \\\hline
      \inc                         & 98                              & 34.10                           & 34.80                           & 55.36                           & 4                     & 33.41                           & 31.63                           & 56.15                           & 4                     & 67.59                          & 68.90                         & {162.02}                           & 14         & \cmark\TP \\\hline
      \inc                         & 66                              & 0.54                            & 0.65                            & 0.64                            & 1                     & 0.32                            & 0.29                            & 0.55                            & 1                     & 1.02                           & 0.80                          & 0.77                               & 1          & \cmark\TP \\\hline
      \inc                         & 31                              & 0.47                            & 0.52                            & 0.48                            & 1                     & 0.21                            & 0.22                            & 0.50                            & 1                     & 0.87                           & 0.66                          & 0.53                               & 4          & \cmark\TP \\\hline
      \inc                         & 31                              & 1.64                            & 1.64                            & 2.26                            & 0                     & 1.23                            & 1.26                            & 2.65                            & 0                     & 4.68                           & 2.69                          & 2.22                               & 3          & \cmark\TP \\\hline
      \inc                         & 80                              & 12.09                           & 11.44                           & 11.93                           & \green{0}             & 10.68                           & 10.33                           & 11.86                           & \green{0}             & 29.25                          & 19.44                         & 16.38                              & {0}        & ---\TN    \\
      \cline{1-2}\cline{4-5}\cline{6-7}\cline{10-11}\cline{14-15}
      \hhline{~~*{1}{=}*{2}{=}~*{3}{=}~*{3}{=}}
      \multicolumn{1}{c}{}         & \multicolumn{1}{c|}{}           & 872.97                          & 883.56                          & {754.74}                        & \multicolumn{1}{c|}{} & 933.78                          & 932.97                          & {807.56}                        & \multicolumn{1}{c|}{} & 1,354.48                       & 1,281.42                      & {1,055.85}
                                   & \multicolumn{1}{c}{}                                                                                                                                                                                                                                                                                                                                                                                                       \\
      \cline{3-5}\cline{7-9}\cline{11-13}
    \end{tabular}
    \begin{tablenotes}
      \ifccstemplate\small\else\footnotesize\fi
      \item
      All times are in seconds.
      $^{\ddag}$ \texttt{basic}~\cite{nuz} and \texttt{symba}~\cite{symba} algorithms are part of \texttt{Z3} whereas \texttt{obvbs} is provided by \texttt{OptiMathSAT}~\cite{OptiMathSAT}.
      $^{\ast}$ \texttt{par.}: parallel bitvector solving with a pool of 8 cores in \texttt{Z3}; \texttt{bp}: encoding $\omega$-classes as Pseudo Boolean equalities in \texttt{Z3};
      \texttt{inc.} incremental bitvector solving via \texttt{CVC5}. \TO: Time-out. \TP: True Positive. \TN: True Negative. \FP: False Positive. \FN: False Negative. \cmark: TVLA detects leakage ($\mathrm{t\mbox{-}test} \ge 10$). ---:~TVLA does not detect any leakage ($\mathrm{t\mbox{-}test} < 10$).

    \end{tablenotes}
  \end{threeparttable}
\end{table*}

\subsubsection*{Detection of New Vulnerabilities}\label{sec:new-vulnerabilities}
\ferhat{
  We want to highlight that the objective of \tool is to find Points of Interest, i.e. potentially vulnerable instruction addresses, so developers can fix them.
  In our work, we used known vulnerabilities to show that \tool can find real vulnerabilities. It can also detect new potential Points of Interest.
  For example, here we report some PoIs in the Post Quantum and Lightweight Crypto libraries as well as constant-time libraries of mbedTLS,
  which can be found in Table~\ref{tab:evaluation}. There are Points of Interest in $\B{15}$, $\B{17}$, $\B{21}$, and $\B{22}$. They are all confirmed with specific t-tests (TVLA) by attacking those PoIs reported by~\tool and using test vectors that \tool generated. We used ChipWhisperer UFO~\cite{2014:204} with STM32F3 target board having 32-bit ARM Cortex-M4 processor core.
}

\revision{
As an example, Listing~\ref{lst:mbedtls} and Listing~\ref{lst:mbedTLS-O3} shows the detected PoI in \B{21} of mbedTLS~\cite{mbedtls}.
The value of \texttt{ret} at line 7 in C code is defined as a sensitive bit-dependent determiner. Its state corresponds to the register \texttt{r0} in the assembly at line 7. Power consumption traces can be classified into two sets depending on the Hamming weight of its value, as the number of cases is 2.
}~\rebuttal{\#7}

\begin{lstlisting}[language=c, caption={\revision{$\text{P}_{21}$: Constant-flow LT comparison.}~\rebuttal{\#7}}, label={lst:mbedtls}, firstline=3, lastline=11, numbers=left, linebackgroundcolor={\ifnum\value{lstnumber}=7\color{light-gray}\fi}]
unsigned mbedtls_ct_mpi_uint_lt(const uint64_t x, 
                                const uint64_t y) {
    uint64_t ret, cond;
    cond = (x ^ y);
    ret = (x - y) & ~cond;
    ret |= y & cond;
    ret = ret >> (sizeof(uint64_t) * 8 - 1);
    return (unsigned)ret;
}
\end{lstlisting}

\begin{lstlisting}[language={[arm]Assembler}, caption={\revision{Disassembly at \texttt{-O3} of Listing~\ref{lst:mbedtls}}~\rebuttal{\#7}}, numbers=left, label={lst:mbedTLS-O3}, firstline=1, lastline=8, linebackgroundcolor={\ifnum\value{lstnumber}=7\color{light-gray}\fi}]
cmp r0, r2       
sbc r2, r1, r3  ; $\uparrow\red{\Delta_\omega} = 32\;\circ\downarrow\red{\Delta_\omega} = 0\phantom{0}\;\Vert\;\red{\tilde{\eta}} = 3.547$ 
eor r0, r1, r3  ; $\uparrow\red{\Delta_\omega} = 32\;\circ\downarrow\red{\Delta_\omega} = 0\phantom{0}\;\Vert\;\red{\tilde{\eta}} = 3.547$
eor r3, r3, r2  ; $\uparrow\red{\Delta_\omega} = 32\;\circ\downarrow\red{\Delta_\omega} = 0\phantom{0}\;\Vert\;\red{\tilde{\eta}} = 3.547$ 
and r0, r0, r3  ; $\uparrow\red{\Delta_\omega} = 32\;\circ\downarrow\red{\Delta_\omega} = 0\phantom{0}\;\Vert\;\red{\tilde{\eta}} = 3.547$ 
eor r0, r0, r2  ; $\uparrow\red{\Delta_\omega} = 32\;\circ\downarrow\red{\Delta_\omega} = 0\phantom{0}\;\Vert\;\red{\tilde{\eta}} = 3.547$
lsr r0, r0, 0x1f; $\uparrow\red{\Delta_\omega} = 1\phantom{0}\;\circ\downarrow\red{\Delta_\omega} = 0\phantom{0}\;\Vert\;\red{\tilde{\eta}} = 0.195\;\;\red{\star}$
bx lr
\end{lstlisting}

\subsubsection*{Performance Evaluation}\label{peformance}
The evaluation is performed on 11th Gen Intel Core™ i7-1185G7 @ 3.00GHz x 8 cores with 32 GB RAM on Ubuntu 20.04.4 LTS OS. \tool's implementation is sequential, however, in the future, we aim to discharge optimization queries concurrently.
In our evaluation we evaluated 24 functions. If a benchmark has loops, we selectively unroll them 8 times. Analyzing benchmarks with 6 different methods take about 2.5 hours to run on the machine. Table~\ref{tab:evaluation} presents the results of the analysis. All times are given in seconds.

The SMT problems that require the capability of finding models that are optimal with regard to some objective functions are grouped
under the umbrella term of Optimization Modulo Theories~\cite{OptiMathSAT:CAV}. In \tool, {differential Hamming weight} and
  {Hamming distance} models specifically requires \emph{single-objective linear optimization} over bitvector terms.

\begin{figure}[h]
  \centering
  \includegraphics[trim={0cm 0cm 1cm 0.19cm}, clip, width=.85\linewidth]{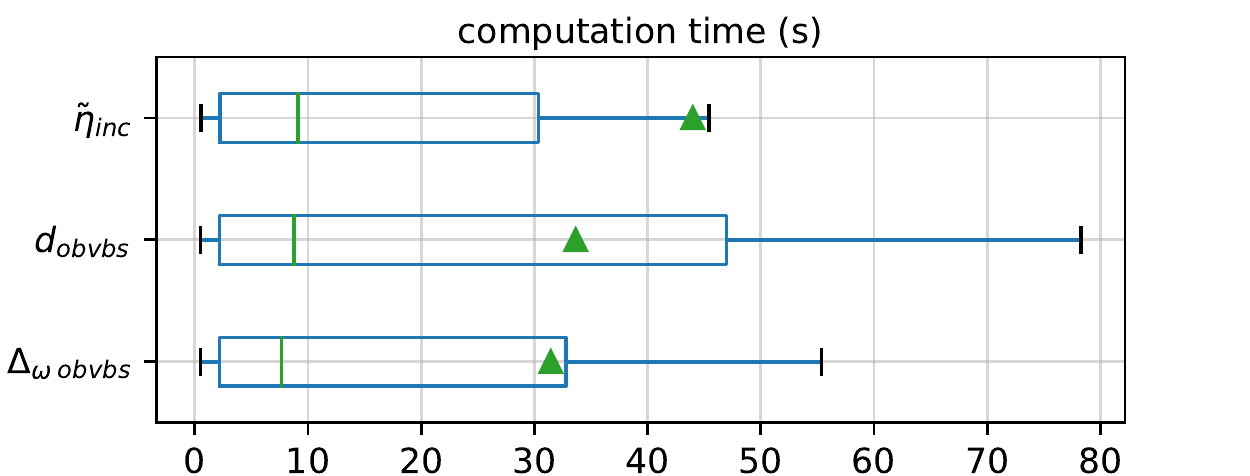}
  \caption{Distributions of computational times show that although median of those methods are close, in average (\lightgreen{$\blacktriangle$})~$\Delta_\omega$ analysis with \texttt{obvbs} performs slightly better.}
  \label{fig:distributions}
\end{figure}

There are efficient SMT-based optimization algorithms in the theory of fixed-size bitvectors:
\texttt{$\nu Z$}~\cite{z3opt} and \texttt{Symba}~\cite{symba} integrated with \texttt{Z3} SMT solver,
and \texttt{OptiMathSAT}~\cite{OptiMathSAT} that extends \texttt{MathSAT5}~\cite{mathsat5} SMT solver.
We incrementally use \emph{bit-vector optimization with binary search} (\texttt{obvbs})~\cite{OptiMathSAT:CAV} algorithm in \texttt{OptiMathSAT}.
In \texttt{$Z3$}, we employ both \texttt{basic}~\cite{nuz} and \texttt{symba}~\cite{symba} optimization methods that
locally enumerate optimal assignments until fixed point. Those methods are able to find intermediate solution along the way.
Therefore, in our experiments, \tool retrieves the latest suboptimal solution in case it runs out of time.
In our work, we compare those three algorithms and found that \texttt{OptiMathSAT}'s \texttt{obvbs} optimization algorithm outperforms \texttt{Z3}'s \texttt{basic} and \texttt{symba} algorithms over our benchmarks (see Figure~\ref{fig:distributions}).
The analysis time varies based on the loop and the complexity of the symbolic expression that represents the destination registers and
the number of discharges on solvers. For instance, in $\B{16}$, the analysis takes 382.49 seconds with Hamming weight model using \texttt{symba} while the instruction count is 149,
whereas, in $\B{17}$, the analysis takes 5.56 seconds with the same method while the instruction count is 223.

With a soft timeout of $60$ seconds given per SMT query, \texttt{Z3} and \texttt{OptiMathSAT} solvers run out of time and returns \emph{suboptimal} solutions for a minimization objective of program $\B{14}$ under Hamming distance model. We couldn't identify what makes this specific query hard for solvers.

In $\omega$-class sampling method, we used \texttt{Z3} solver to solve bitvector queries in parallel mode with a pool of 8 cores (see \texttt{par.} in Table~\ref{tab:evaluation}) and compare the performance with two other methods: encoding $\omega$-class constraints as Pseudo-Boolean equalities in \texttt{Z3} (\texttt{pb.}) and incremental encoding using \texttt{CVC5}~\cite{cvc5} (\texttt{inc.}).
We also set full timeout for algorithms, which is $600$ seconds per benchmark and method. \texttt{par.} and \texttt{pb.} timed out for benchmark $\text{P}_{16}$ and thus
\texttt{inc.} method on \texttt{CVC5} solver outperforms \texttt{par.} and \texttt{pb.} methods, however, if we remove this benchmark, method \texttt{pb.} outperforms others.

\subsubsection*{Randomized and Shuffled Implementations}\label{sec:quantification}
\setcounter{program}{6} \inc, \inc, \inc, \inc, and \inc~are different protected implementations of benchmark $\B{6}$ which is NIST's PQC Standardization winner CRYSTALS-Kyber's vulnureable message encoding function.
\revision{
  The vulnureability that leads to the attack path is quite similar to that of New Hope's message encoding. The message $msg$ is processed in bitwise manner during the message encoding as shown in Listing~\ref{lst:kyber}.
  The mask value can be either $0x0000$ or $0xFFFF$; therefore, the number of cases of the mask value is 2 (Definition~\ref{def:determiner}). Moreover, the differential Hamming weight is equal to $\Delta_\omega = 16$. \cite{2020:992} exercised single-trace attacks at this PoI. More recently \cite{cryptoeprint:2023/294} and \cite{cryptoeprint:2022/1713} demonstrated attacks on its masked version~\cite{cryptoeprint:2022/058}.
}\rebuttal{\#8}

\begin{lstlisting}[language=c, caption={\revision{$\text{P}_{6}$: CRYSTALS-Kyber's message encoding}}, label={lst:kyber}, firstline=23, lastline=34, numbers=left, linebackgroundcolor={\ifnum\value{lstnumber}=9\color{light-gray}\fi}]
void poly_frommsg(poly *r, 
      const uint8_t msg[KYBER_INDCPA_MSGBYTES]) {
  unsigned int i, j;
  int16_t mask;
  for (i = 0; i < KYBER_N / 8; i++) {
   for (j = 0; j < 8; j++) {
     mask = -(int16_t)((msg[i] >> j) & 1);
     /*$\uparrow\red{\Delta_\omega} = 16\;\circ \downarrow\red{\Delta_\omega} = 16\;\Vert\uparrow\red{\Delta_d} = 16\;\circ\downarrow\red{\Delta_d} = 16\;\Vert\;\red{\tilde{\eta}} = 1.00$*/
     r->coeffs[8*i + j] = mask & ((KYBER_Q+1)/2);
    }
  }
}
\end{lstlisting}

\newcommand{\cImp}{\Gape[-2pt]{\thead{\bfseries{CRYSTALS-Kyber Message Encoding}\\\bfseries{converts 32-byte message to polynomial}}}}
\newcommand{\cSuccess}{\Gape[-2pt]{\thead{\bfseries{Attack}\\\bfseries{Success}}}}
\newcommand{\cResult}{\Gape[-2pt]{\thead{\bfseries{\green{Pascal}}\\\bfseries{detects?}}}}
\newcommand{\cTtest}{\Gape[-2pt]{\thead{\bfseries{t-test}\\\bfseries{$t_{max}$}}}}
\newcommand{\cPrecision}{\Gape[-2pt]{\thead{\bfseries{Pascal's}\\\bfseries{Precision}}}}
\setcounter{program}{5}
\begin{table}[ht!]
  \setlength{\tabcolsep}{0.25em}
  \ifccstemplate\small\else\footnotesize\fi
  \caption{\tool's precision of vulnerability detection for {Shuffling} and {Randomization} countermeasures~\cite{2021:1307} applied to NIST's PQC Standardization winner CRYSTALS-Kyber's message encoding function.}
  \label{tab:evaluation-against-kyber-countermeasure}
  \centering
  \begin{tabular}{@{}l|l|cc|c@{}}
    \toprule 
         & \cImp                                                  & \cTtest & \cSuccess & \bfseries{Pascal} \\
    \midrule
    \inc & Unprotected NIST Submission \cite{2021:1307, 2020:549} & 437     & 100\%     & \TP               \\
    \midrule
    \inc & Message encoding with multiplication \cite{2020:368}   & 177.0   & 100\%     & \TP               \\
    \inc & Data independent polynomial gen. \cite{2021:1307}      & 24.8    & 68.6\%    & \TP               \\
    \inc & Balanced data independent poly. gen. \cite{2021:1307}  & 19.6    & 67.9\%    & \TP               \\
    \inc & Polynomial randomization \cite{2021:1307}              & 13.8    & 64.0\%    & \TP               \\
    \inc & Byte and bit level random ordering \cite{2021:1307}    & 5.2     & 50.1\%    & \FP               \\
    \bottomrule
  \end{tabular}
\end{table}

Table~\ref{tab:evaluation-against-kyber-countermeasure} shows \tool's precision in comparison with t-test measurements and success rates of power analysis attacks based on the work of \cite{2021:1307}.
A t-test result below 10 indicates absence of a leakage. Individual shuffling and masking countermeasures were shown to be vulnerable against simpler power attacks, and \tool successfully confirmed vulnerabilities in $\B{7}$, $\B{8}$, $\B{9}$ and $\B{10}$.
A combination of masking and shuffling increases the trace requirement for the attack, and benchmark $\B{11}$ shuffles message encoding using byte and bit level random ordering. If we compare it with the reference implementation, $\B{5}$, this countermeasure significantly reduces the success rate of attacks from 100\% to 50.1\% while introducing 2.49X overhead~\cite{2021:1307}. In this benchmark, \tool marked an instruction address as vulnerable (False Positive). However, once the generated report is investigated, one can realize that the vulnerability manifests itself randomly at some loop iterations, and therefore, it can be corrected by the user as True Negative.
\revision{
  $\B{11}$ is detailed in Listing~\ref{lst:kyber-countermeasure} and shows `Data independent polynomial generation with balanced byte look-up' countermeasure proposed by \cite{2021:1307} for Kyber's message encoding function (Listing~\ref{lst:kyber}). 
  The strategy is to shift the pointer array \texttt{p\_r} and the balancing array \texttt{xorMask} by $0$ or $1$, depending on the most significant bit of the first message byte (MSB). As the message \texttt{msg} is randomly chosen, evaluating the MSB serves as a source of randomness without introducing an additional fetch from a random number generator.
}\rebuttal{\#8}

\begin{lstlisting}[language=c, caption={\revision{$\text{P}_{11}$: Protected version of Listing~\ref{lst:kyber}.~\rebuttal{\#8}}}, label={lst:kyber-countermeasure}, firstline=24, lastline=59, numbers=left, linebackgroundcolor={\ifnum\value{lstnumber}=30\color{light-gray}\else\ifnum\value{lstnumber}=31\color{light-gray}\fi\fi}]
void $\red{masked}$_poly_frommsg(poly *r, 
    const uint8_t msg[KYBER_INDCPA_MSGBYTES]) {
  unsigned int i, j;
  poly r_d;
  poly *p_r[256 + 1];
  uint32_t xorMasks[3] = {0xaaaaaaaa, 0x55555555, 
    0xaaaaaaaa};
  for (i = 0; i < 256; i += 2) {
    p_r[i] = r; 
    p_r[i + 1] = &r_d; 
  }
  for (i = 0; i < KYBER_N; i++) {
    r->coeffs[i] = (KYBER_Q + 1)/2;
    r_d.coeffs[i] = (KYBER_Q + 1)/2; 
  }
  uint32_t rand = (0xaaaa00aa ^ msg[0]);
  uint8_t i_m = rand & 0x1f;
  uint8_t j_m = (rand >> 5) & 0xff;
  uint32_t b_inv = (rand >> 7) & 0xff;
  for (i = 0; i < 255; i++) {
    *(p_r + i) = *(p_r + i + b_inv); }
  for (i = 0; i < 2; i++) {
    *(xorMasks + i) = *(xorMasks + i + b_inv); 
  }
  uint8_t i_r, j_r;
  for (i = 0; i < KYBER_N/8; i++) {
    i_r = i ^ i_m;
    for (j = 0; j < 8; j++) {
      j_r = j ^ j_m;
      p_r[((xorMasks[(j_r & 1)]^msg[i_r]) >> j_r) 
        & 0xff]->coeffs[8 * i + j] = 0; }
  }
}
\end{lstlisting}

We expect no False Negatives since if there is some Hamming weight difference or significant entropy loss, \tool will detect such instructions and flag their addresses as Point of Interests. \tool only reports a False Negative in \lstinline|poly_Z3_to_Zq| (benchmark $\B3$) unless the precondition on input variable is given i.e. the input coefficients must be one of $\{0, 1, 2\}$.

In benchmarks $\B{5}$, $\B{15}$, and $\B{24}$, neither of the models detect any vulnerabilities and in fact we weren't able to find a significant t-test value above $10$. Benchmark $\B{23}$, ARX box construction of SPARX~\cite{sparx} (see Listing~\ref{lst:arxbox-arm}), is vulnerable while benchmark $\B{24}$, ARX box of NIST's LWC finalist Sparkle~\cite{sparkle} is secure; \tool did not detect any PoI, and we did not detect any presence of power leakage via TVLA.

\revision{
\subsubsection*{Target Architectures}
\label{sec:target-architectures}
We have confirmed the presence of those vulnerable PoIs listed in Table~\ref{tab:vulnurebilities} using TVLA on ARM Cortex-M4 since it is widely used in the PQC community~\cite{pqm4:repo, pqm4}. We also separately confirmed the applicability of the Hamming weight leakage model to single-trace side-channels on different architectures: ARM Cortex-M4, ARM Cortex-M0, Atmel AVR XMEGA, and MSP430 using STM32F3/F4, STM32F0, ATXmega128D4-AU, and MSP430FR5994 targets respectively.~\rebuttal{\#1.b}

Any circuit not explicitly designed to be resistant to power attacks exhibits data-dependent power consumption phenomenon~(cf. Section~\ref{sec:power-analysis}), and Hamming weight leakage model is a well-established method in the literature to model this behavior. Instruction-level information that is obtained from the disassembler retains enough symbolic information to track data-dependent Hamming-weight characteristics.~\rebuttal{\#1.a}

The Hamming weight leakage model has been used in recent power analysis attacks on modern Intel (and AMD) x86 CPUs, as demonstrated in the Platypus~\cite{platypus} and Hertzbleed~\cite{hertzbleed} attacks. In the Hertzbleed attack, the vulnerability on {SIKE}~\cite{sike}'s {\itshape three point ladder} is investigated using the Hamming weight leakage model and it is also detected by \tool with symbolical register analysis. Additionally, in the Platypus work, the performance counters of x86 are used and attacks are performed using the Hamming weight leakage model.
}~\rebuttal{\#1.b}

\subsubsection*{Discussions}
\label{sec:discussions}
We argue that our work highlights the need for single-trace side-channel aware constant-time cryptographic coding.
The conflict between a constant-time implementations of cryptographic algorithms and its single-trace power or EM leakages manifested in hardware makes this a non-trivial task and should be investigated.

Experimenting with power side-channel attacks becomes affordable~\cite{hw-hack-book} thanks to special power side-channel analysis circuitry such as ChipWhisperer~\cite{2014:204}. Additionally, protecting against power analysis attacks is less straightforward and usually more expensive than countermeasures for timing attacks.
Simple countermeasures such as introducing jitter adds horizontal noise leading to non-alignment of PoI across measurements, and it increases the attack effort.

\revision{
  Our current technique is able to detect multi-trace vulnurebilities, as shown in the Listing~\ref{lst:arxbox} where the attacker needs to differentiate four different values. However, in this paper, we have primarily focused on single-trace attacks as they are the simplest form of attack for an attacker to perform on unprotected implementations.
}\rebuttal{\#4}

\section{Related Work}
\label{sec:related-work}

\tool's approach to detection to power/EM side-channels vulnerabilities has a unique position in the literature since it is the first to devise a leakage detection technique using automated reasoning according to the SoK study conducted in \cite{2021:497} (see Table~\ref{tab:related-work} for a summary). \tool does not use any simulated, estimated, or measured trace and covers both power and EM side-channels; it does not require any microarchitectural information and therefore it is not tailored to any specific architecture, and it is publicly available. Furthermore, it approximately quantifies side-channel vulnerabilities. However, it is not the only method to detect power side-channel vulnerabilities. In this section, we discuss the most relevant tools in the literature.

\begin{table}
  \caption{Existing Power/EM Side Channel Analysis tools and methods, adapted from the SoK paper of~\cite{2021:497}. Other than~\tool, all tools are leakage simulators.}
  \label{tab:related-work}
  \setlength{\tabcolsep}{0.2em}
  \footnotesize
  \centering
  \begin{threeparttable}
    \begin{tabular}{@{}llcllcc@{}}
      \toprule 
      Work                     & Year & LM$^\ddagger$   & Target Device(s)            & SC               & Avail.$^\|$ \\
      \midrule
      \bf{\tool}               & 2022 & \harveyBallNone & \green{Any Target}$^{\dag}$ & \green{Power/EM} & \gcmark     \\
      Rosita++~\cite{rosita++} & 2021 & \harveyBallHalf & ARM Cortex M0/M4            & Power            & \cmark      \\
      Emsim~\cite{emsim}       & 2020 & \harveyBallHalf & Risc-V                      & EM               & \xmark      \\
      Rosita~\cite{rosita}     & 2019 & \harveyBallHalf & ARM Cortex M0               & Power            & \cmark      \\
      Elmo~\cite{elmo:1}       & 2017 & \harveyBallHalf & ARM Cortex M0               & Power            & \cmark      \\
      Ascold~\cite{ascold}     & 2017 & \harveyBallHalf & ATMega163                   & ILA              & \cmark      \\
      Savrasca~\cite{savrasca} & 2017 & \harveyBallFull & ATMega163                   & Power            & \cmark      \\
      \cite{reparaz}           & 2016 & \harveyBallFull & not relevant                & Power            & \xmark      \\
      Sleak~\cite{sleak}       & 2014 & \harveyBallFull & ARM Cortex A8               & Register         & \xmark      \\
      Silk~\cite{silk}         & 2014 & \harveyBallFull & ATmega328P                  & Power            & \cmark      \\
      \cite{gagnerot}          & 2013 & \harveyBallFull & Risc-V                      & Power            & \xmark      \\
      \cite{debande}           & 2012 & \harveyBallHalf & not specified               & Power            & \xmark      \\
      \cite{oscar}             & 2009 & \harveyBallFull & AT90XX, ATmegaXX            & Power            & \xmark      \\
      \cite{riscure:sca}       & 2002 & \harveyBallFull & not relevant                & Power            & \xmark      \\
      Pinpas~\cite{pinpas}     & 2003 & \harveyBallFull & smartcards                  & Power            & \xmark      \\
      \bottomrule
    \end{tabular}
    \begin{tablenotes}
      \ifccstemplate\small\else\footnotesize\fi
      \item ${\dag}$ Hamming weight Leakage assumption is verified on these targets: ATXmega128D4-AU, STM32F3/F4 (ARM Cortex-M4), STM32F0 (ARM Cortex-M0), MSP430FR5994. $\ddagger$ Leakage Model:~\harveyBallFull: a black-box model (ISA level information needed);~\harveyBallHalf: a gray-box model (microarchitectural information needed); and \harveyBallNone: a white-box model (formal instruction semantics needed). $^\|$: publicly available or not. \\
    \end{tablenotes}
  \end{threeparttable}
\end{table}

The Test Vector Leakage Assessment (TVLA)~\cite{tvla} identifies differences between two sets of side channel measurements by computing the t-test for the two sets of measurements. It is being used in the literature to confirm the {\itshape presence} or {\itshape absence} of side leakages for power traces. TVLA does not pinpoint code locations, only statistically says if there may be some leakage. It also requires partitioning of the traces based on the value of a particular bit of an intermediate state in the targeted algorithm, and therefore to comprehensively evaluate an implementation, single bit of every single intermediate state must be tested~\cite{2013:298}, but this is impractical, the analysis are usually restricted to measure not more than one million traces~\cite{2015:1215}. 
We would like to clarify that the goal of \tool is to automatically find potentially vulnerable instructions. Pascal is not a substitute for, or an alternative to, methods such as TVLA. Once a Point of Interest is found automatically, a developer can then investigate it, possibly using TVLA to confirm if this is indeed a vulnerability and fix it.

There are leakage emulators in the literature such as Elm0~\cite{elmo:1}, Rosita~\cite{rosita}, and Rosita++~\cite{rosita++} for ARM Cortext-M0/M4, Ascold~\cite{ascold} for ATMega163 targets. 
These tools are able to simulate instruction-based power variations on specific target architectures based on profiling or reverse engineering a target device, and generate emulated traces.
However, creating such a model is prohibitively effort-intensive, even for relatively simple processors (in-order, no cache)~\cite{2021:497}.  
They are not able to formally guarantee their existence as \tool does. 
On the other hand, those leakage simulators may identify power variations due to the underlying microarchitecture that might not be explained by Hamming weight or Hamming distance leakage models. However, all the attacks we investigated so far can be accounted for by Hamming weight models. 
\section{Conclusions}
\label{sec:conclusions}

To help make the process of detecting potential power side-channel vulnerabilities easier for cryptographers, this work presented Pascal that introduces a number of novel symbolic register analysis techniques in binary analysis of constant-time cryptographic implementations, and pinpoints locations of potential power side-channel vulnerabilities with high precision. It also generates test vectors for TVLA analysis. Our tool was able to locate all currently reported single-trace power side-channel vulnerabilities in constant-time code of post-quantum cryptographic algorithms. The analysis of target functions can be done automatically with Pascal, and significantly reduces the burden on the programmer for finding potentially vulnerable code locations.

\bibliographystyle{plain}
\bibliography{IEEEabrv, pascal}

\appendix

\begin{figure*}[t]
  \includegraphics[trim={0.2cm 2cm 0.1cm 0.2cm}, clip, width=\linewidth]{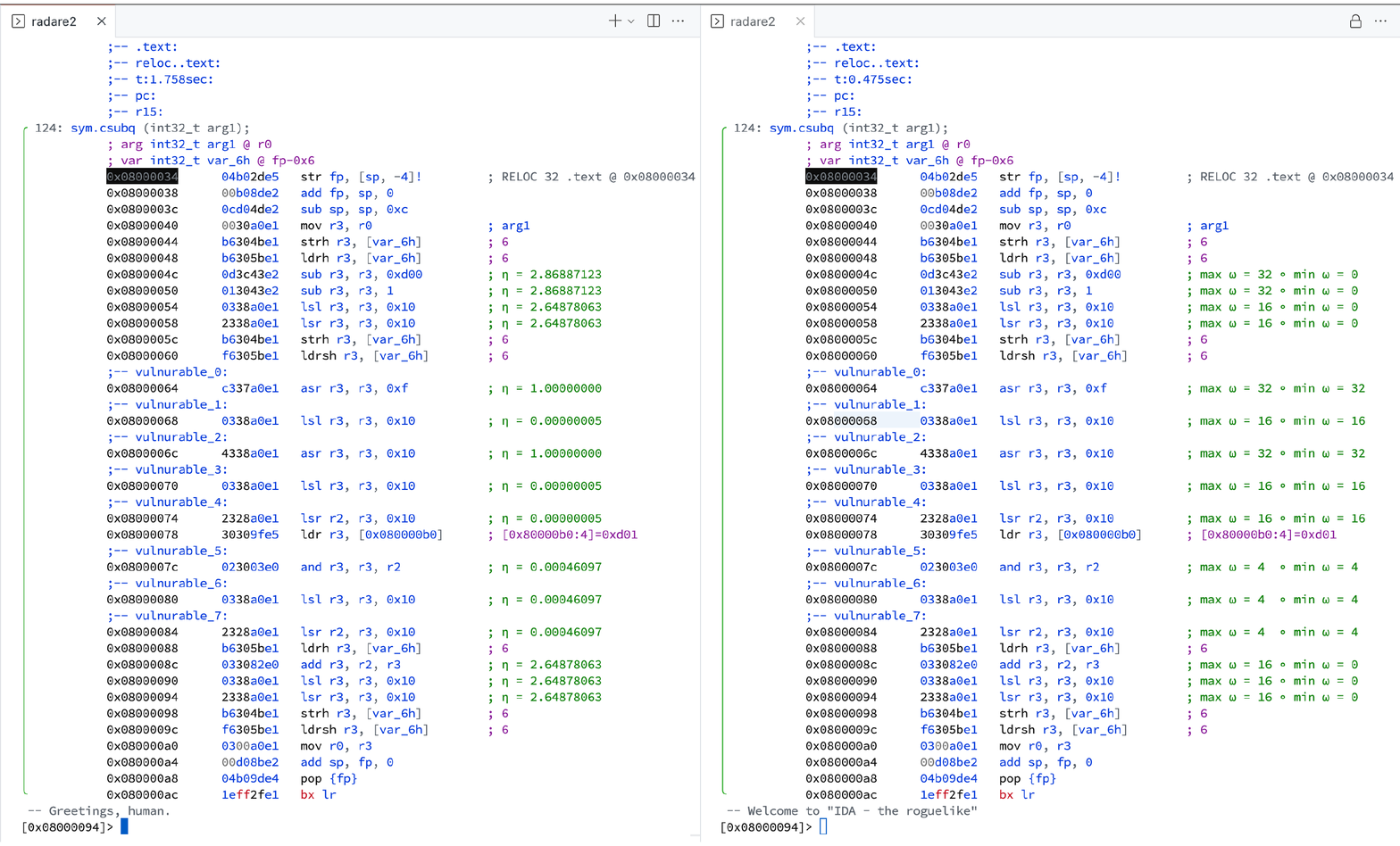}
  \caption{\tool's Radare2 output for Kyber's conditional subtraction function ---csubq--- each secret-tainted arithmetic/logical instructions are annotated with Min and Max Hamming weights and approximate entropy. Point of Interests in Single-Trace side-channel attacks are flagged as \emph{vulnerable}.} \label{fig:kyber-csubq}
  \ifccstemplate\Description{}\fi
\end{figure*}

\subsection{Discussion of Existing Vulnerabilities}\label{sec:discussion-of-existing-vulnerabilities}

\begin{table*}[ht!]
  \caption{Single-trace Power/EM side-channel attacks that our tool confirms}
  \label{tab:attack-classification}
  \footnotesize
  \setlength{\tabcolsep}{0.8em}
  \centering
  \begin{threeparttable}
    \begin{tabularx}{\linewidth}{@{}lllllllllX@{}}
      \toprule
      \bf Work          & \bf Year & \bf Type & \bf Crypto Algorithm  & \bf Operation             & \bf Method & \bf Classifier & \bf Side-Channel Attack Type            \\
      \midrule
      \cite{2021:HOST}  & 2021     & Power    & NTRU, Dilithium       & Polynomial Sampling       & SOSD       & NPDF           & Single-Trace Template-based Attack      \\
      \cite{2021:1307}  & 2021     & Power    & Crystals-Kyber        & Message Encoding          & SOST       & t-test         & Single-Trace Simple Power Attack        \\
      \cite{2021:790}   & 2021     & EM       & NTRU                  & Modular Reduction         & N/A        & RMS            & Single-Trace Simple Power Attack        \\
      \cite{2021:858}   & 2021     & Power    & SIKE                  & Three Point Ladder        & N/A        & PCC            & Single-Trace Correlation Power Analysis \\
      \cite{2021:104}   & 2021     & Power    & Lattice-based         & Masked Comparisons        & TVLA       & t-test         & Single-Trace Simple Power Attack        \\
      \cite{2020:1559}  & 2020     & EM       & LWE/LWR based         & Message Decoding          & TVLA       & t-test         & Single-Trace Template-based Attack      \\
      \cite{2020:992}   & 2020     & Power    & LWE/LWR based         & Message Encoding          & SOST       & ML-based       & Single-Trace Template-based Attack      \\
      \cite{2020:549}   & 2020     & EM       & Lattice-based         & Message Encoding          & TVLA       & t-test         & Single-Trace Template-based Attack      \\
      \cite{2020:368}   & 2020     & Power    & NewHope               & Message Encoding          & N/A        & S              & Single-Trace Simple Power Attack        \\
      \cite{2019:1236}  & 2019     & Power    & HQC                   & Error Correction          & N/A        & k-means        & Single-Trace Template-based Attack      \\
      \cite{2019:948}   & 2019     & EM       & Round5/LAC            & Error Correction          & TVLA       & t-test         & Chosen-Ciphertext Clustering Attack     \\
      \cite{2019:948}   & 2019     & EM       & Lattice-based         & FO Transform              & TVLA       & t-test         & Chosen-Ciphertext Clustering Attack     \\
      \cite{2019:335}   & 2019     & Power    & Sparx                 & ARX-box Assembly          & N/A        & PCC            & Correlation Power Analysis Attack       \\
      \cite{2018:1}     & 2018     & Power    & NTRU                  & Polynomial Multiplication & N/A        & Averaging      & Single-Trace Simple Power Attack        \\
      \cite{sim2018key} & 2018     & EM       & Scalar Multiplication & Key bit check phase       & SOST       & k-means        & Single-Trace Template-based Attack      \\
      \cite{sim2017key} & 2017     & Power    & Scalar Multiplication & Key bit check phase       & SOST       & k-means        & Single-Trace Template-based Attack      \\
      \cite{2016:923}   & 2016     & Power    & Curve25519            & Montgomery Ladder         & N/A        & NPDF           & Single-Trace Template-based Attack      \\
      \bottomrule
    \end{tabularx}
    \begin{tablenotes}
      \ifccstemplate\small\else\footnotesize\fi
      \item{Classifier: Classification Method (Statistical Analysis for Clustering), EM: Electromagnetic Emanation, Power: Power-related side-channel attacks, PoI: Interesting points in time on the traces, TVLA: Test Vector Leakage Assessment~\cite{tvla}, SOSD: Sum of Squared pairwise Differences of the average signals~\cite{mia}, SOST: Sum of Squared pairwise t-differences~\cite{sost}, S: Sum of Squared differences, k-means: k-means clustering algorithm, NPDF: Normal Probability Density Function, PCC: Pearson Correlation Coefficient~\cite{CPA}, RMS: Root Mean Square.}
    \end{tablenotes}
  \end{threeparttable}
\end{table*}

\begin{figure*}[t]
  \centering
  \begin{subfigure}{.99\columnwidth}
    \centering
    \includegraphics[trim={0.5cm 0.0cm 1.7cm 1.2cm}, clip, width=\linewidth]{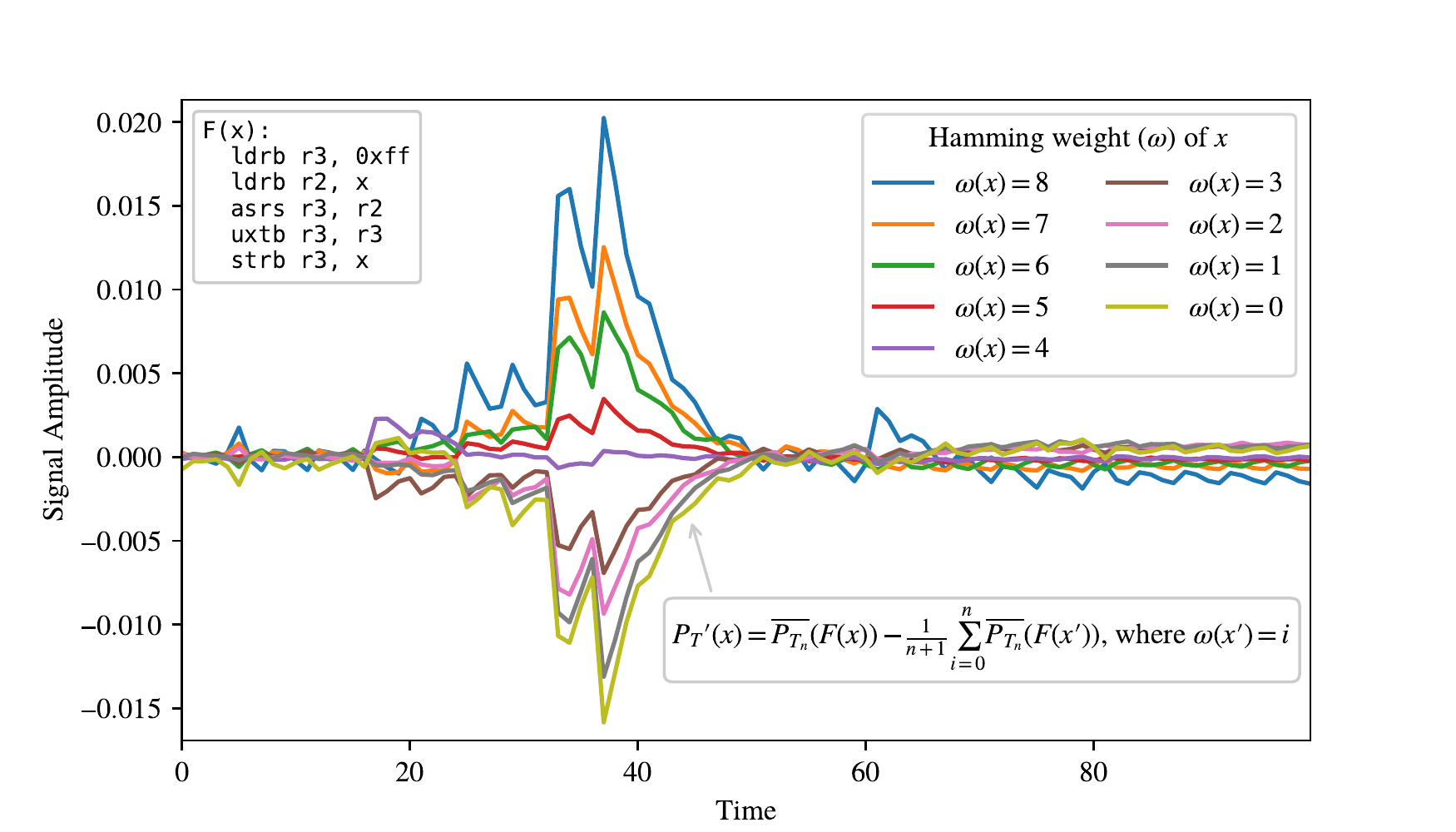}
  \end{subfigure}%
  \qquad
  \begin{subfigure}{.99\columnwidth}
    \centering
    \includegraphics[trim={0.5cm 0.0cm 1.7cm 1.2cm}, clip, width=\linewidth]{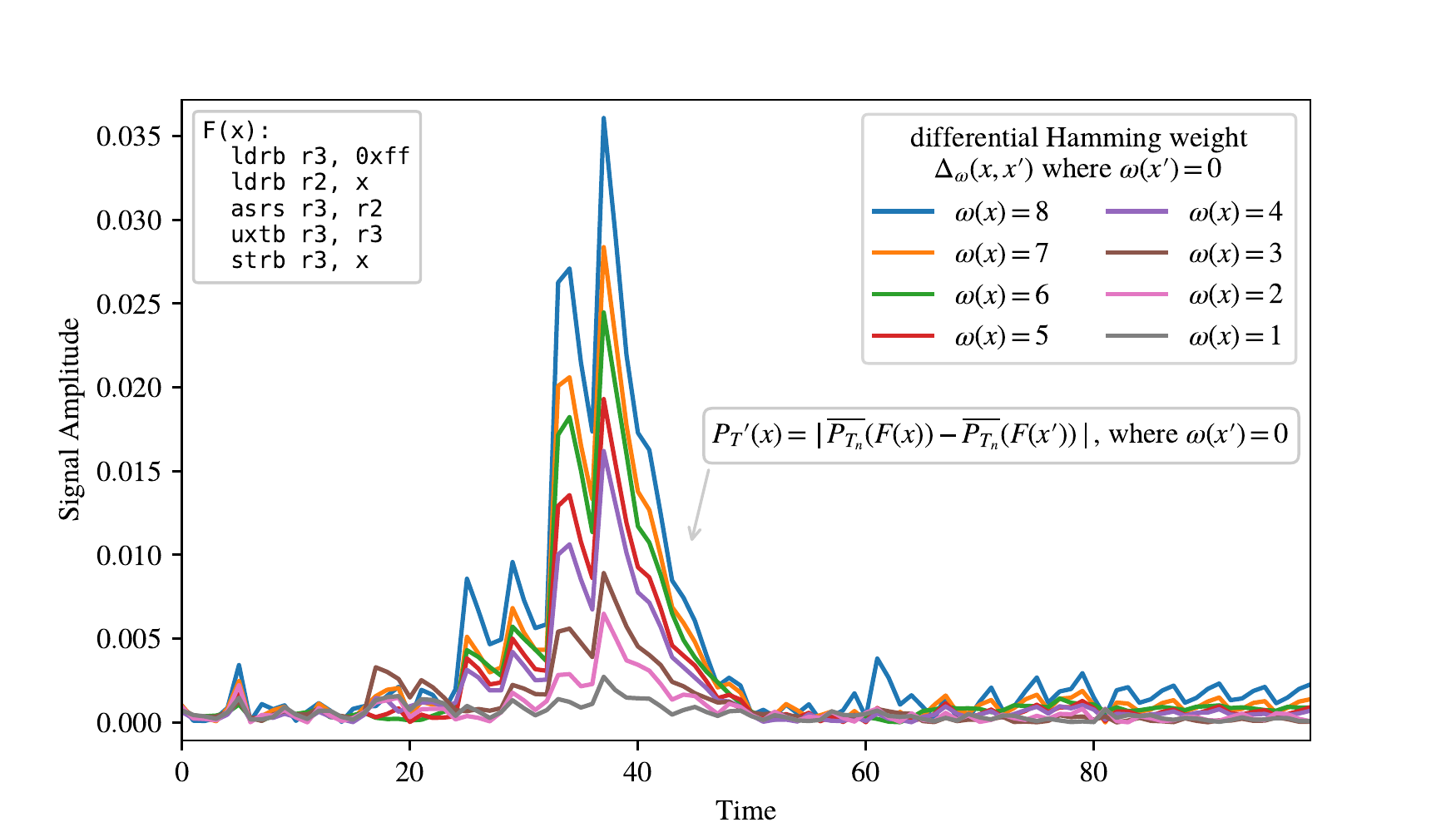}
  \end{subfigure}
  \caption{ Observable difference in power traces. Execution of $R_d := R_d \gg x$ with all $\omega$ classes of $\mathbb{F}_8$. 1000 samples per $\omega$ class collected from a 32-bit ARM Cortex-M4 (STM32F3) target.}
  \label{fig:hw-vs-delta-hw}
\end{figure*}
\begin{figure*}[t]
  \centering
  \begin{subfigure}{\columnwidth}
    \centering
    \includegraphics[trim={0.5cm 0.0cm 1.7cm 1.2cm}, clip, width=\linewidth]{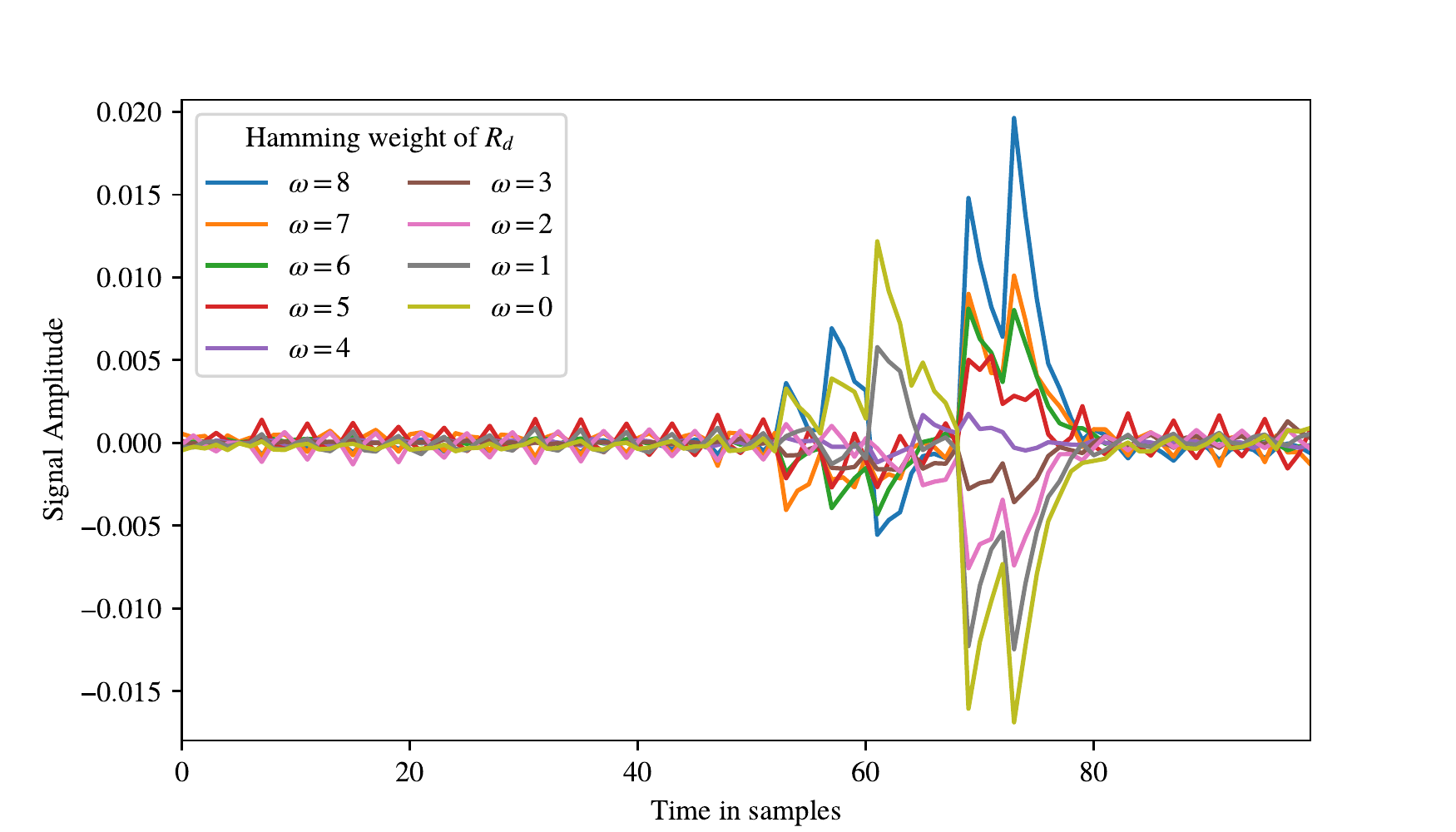}
  \end{subfigure}%
  ~
  \begin{subfigure}{\columnwidth}
    \centering
    \includegraphics[trim={0.5cm 0.0cm 1.7cm 1.2cm}, clip, width=\linewidth]{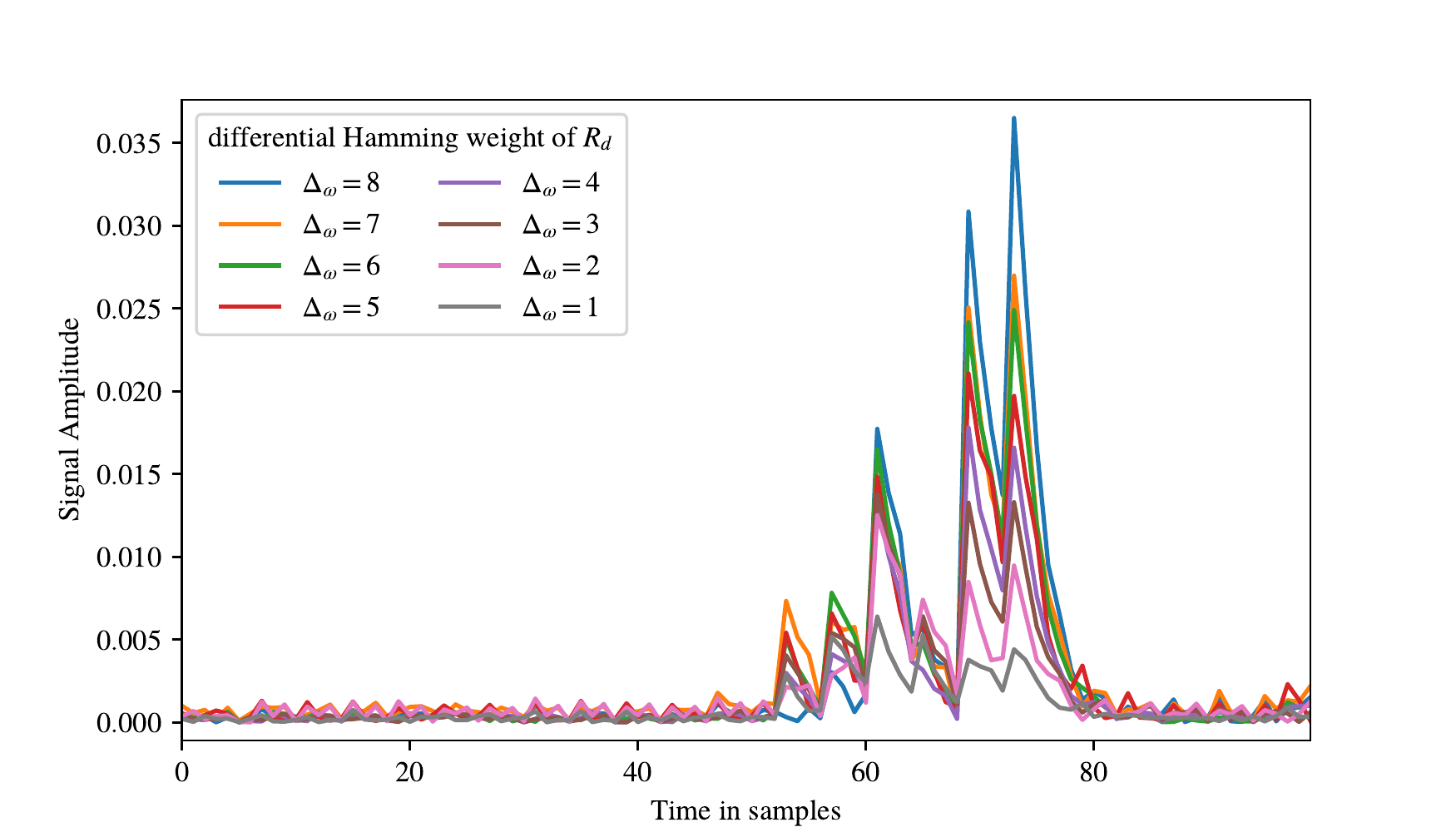}
  \end{subfigure}
  \caption{\revision{Observable difference in power traces. Execution of $R_d := R_d \gg x$ with all $\omega$ classes of $\mathbb{F}_8$. 1000 samples per $\omega$ class collected from a 32-bit ARM Cortex-M0 (STM32F0) target.}~\rebuttal{\#1.b}}
  \label{fig:hw-vs-delta-hw-m0}
  \ifccstemplate\Description{}\fi
\end{figure*}
\begin{figure*}[t]
  \centering
  \begin{subfigure}{0.99\columnwidth}
    \centering
    \includegraphics[trim={0.5cm 0.0cm 1.7cm 1.2cm}, clip, width=\linewidth]{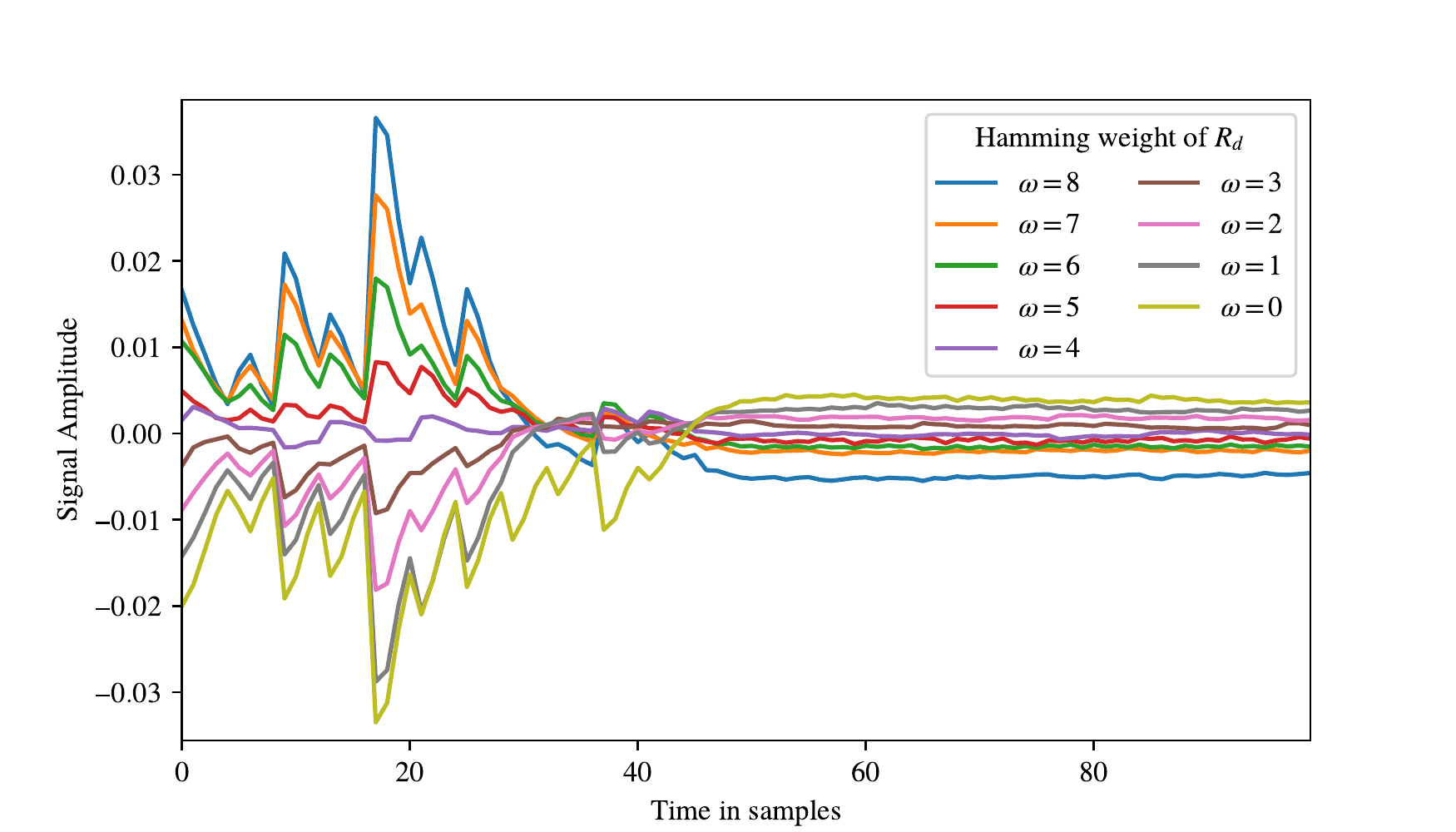}
  \end{subfigure}%
  \qquad
  \begin{subfigure}{0.99\columnwidth}
    \centering
    \includegraphics[trim={0.5cm 0.0cm 1.7cm 1.2cm}, clip, width=\linewidth]{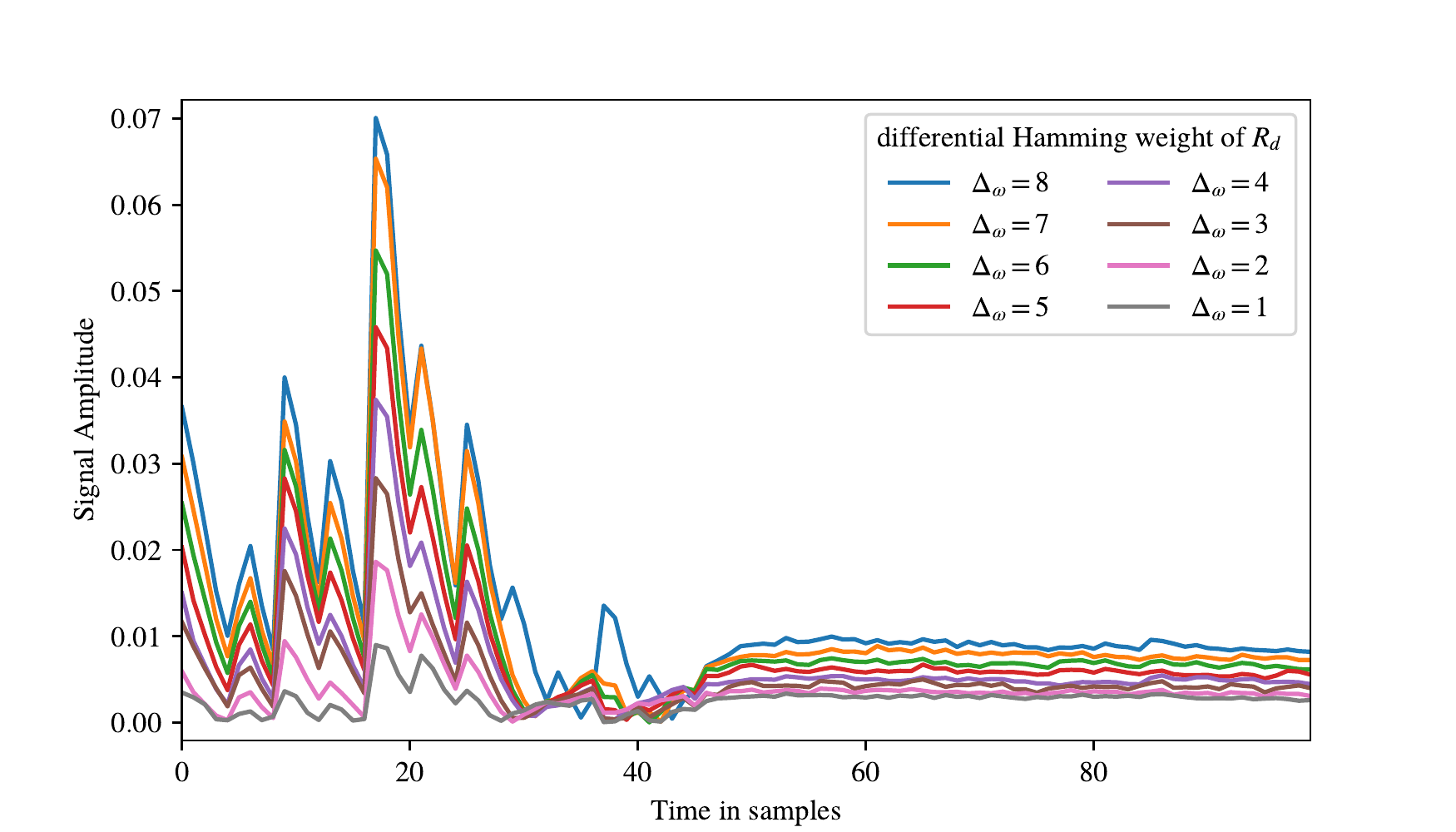}
  \end{subfigure}
  \caption{\revision{Observable difference in power traces. Execution of $R_d := R_d \gg x$ with all $\omega$ classes of $\mathbb{F}_8$. 1000 samples per $\omega$ class collected from a XMEGA 8-bit RISC target.}~\rebuttal{\#1.b}}
  \label{fig:hw-vs-delta-hw-asr-avr}
  \ifccstemplate\Description{}\fi
\end{figure*}

\cite{sim2018key} present a single trace attack based on the power consumption properties of the key bit check of scalar multiplication used in Elliptic Curve Cryptography.  Scalar multiplication and modular exponentiation consist of iterative operations associated with the private key bit $k_i$ value. 
Accordingly, at the beginning of each iteration, the key bit value is extracted from an $n$-bit key string, $k = (k_{n-1}, k_{n-2}, \cdots, k_0)_2$ and stored in a $k_i$ variable. Thus, power consumption is related to the $k_i$ value. In software implementation, power consumption in the key bit check phase is associated with the Hamming weight of $k_i$ $(0 \le i \le n-1)$, i.e., if $k_i = 0$, power consumption related to 0 occurs. Otherwise, power consumption is associated with 1 (power consumption differs when leakage is zero or one, i.e. $P(\omega(0)) \ne P(\omega(1))$). 

Listing~\ref{lst:openssl} shows the key identification function from OpenSSL~\cite{openssl}. The PoI comes immediately after the ``\lstinline[basicstyle=\ttfamily\small]{& ((BN_ULONG)1)}'' operation is performed at the destination register.

\begin{lstlisting}[language=c, caption={Key bit identification function of OpenSSL}, label={lst:openssl}, firstline=11, lastline=19, numbers=left, linebackgroundcolor={\ifnum\value{lstnumber}=8\color{light-gray}\fi}]
int BN_is_bit_set(const BIGNUM *a, int n) {
  int i, j;
  bn_check_top(a);
  if (n < 0) return (0);
  i = n / BN_BITS2;
  j = n % BN_BITS2;
  if (a->top <= i) return (0);
  return (int)(((a->d[i]) >> j) & ((BN_ULONG)1));
}
\end{lstlisting}

Digital signature schemes generate a valid signature on a message using a secret key and the signature's authenticity can be verified with the associated public key. 
CRYSTALS-Dilithium~\cite{2017:633} is a lattice-based digital signature scheme and one of the three finalists running for the NIST's post-quantum digital signature standard. 
An adversary can target the random challenge sampling during the signature generation at line 17 in Listing~\ref{lst:dilithium} where the implementation determines the sign of the non-zero coefficients. 
\begin{lstlisting}[language=c, caption={Dilithium polynomial generation}, label={lst:dilithium}, numbers=left, linebackgroundcolor={\ifnum\value{lstnumber}=17\color{light-gray}\fi}]
void poly_challenge(poly *c, 
            const uint8_t seed[SEEDBYTES]) {
  ... 
  for (i = 0; i < 8; ++i) 
    signs |= (uint64_t)buf[i] << 8 * i;
  pos = 8;
  for (i = 0; i < N; ++i) c->coeffs[i] = 0;
  for (i = N - TAU; i < N; ++i) {
    do {
      if (pos >= SHAKE256_RATE) {
        shake256_squeezeblocks(buf, 1, &state);
        pos = 0;}
      b = buf[pos++];
    } while (b > i);
    c->coeffs[i] = c->coeffs[b];
    /* $\uparrow\red{\Delta_\omega = 31}\;\circ\;\downarrow\red{\Delta_\omega = 31}$ */
    c->coeffs[b] = 1 - 2 * (signs & 1); 
    signs >>= 1;
  }
}
\end{lstlisting}
This operation can leak information about how many negative and positive coefficients the private polynomial has, which gives a hint about the secret challenge. \cite{2021:HOST} target this coefficient assignment operation in Listing~\ref{lst:dilithium} through a single-trace template attack. The assignment possible outcomes are -1 ($0xFFF\ldots F$) with $\omega = 32$ and 1 ($0x000\ldots 1$) with $\omega = 1$; hence, two significantly different power measurements due to the high Hamming weight difference.

\begin{lstlisting}[language=c, caption={NTRU Comparison}, label={lst:ntru}, numbers=left, linebackgroundcolor={\ifnum\value{lstnumber}=6\color{light-gray}\fi}]
#define int32_MINMAX(a, b)                            
do {                                                  
  int32_t ab = (b)^(a);                               
  int32_t c = (int32_t)((int64_t)(b)-(int64_t)(a)); 
  c ^ = ab & (c ^(b));                                
  c >>= 31;  // $\uparrow\red{\Delta_\omega = 32}\;\circ\;\downarrow\red{\Delta_\omega = 32}$                                         
  c &= ab;                                            
  (a) ^ = c;                                          
  (b) ^ = c;                                          
} while (0)
\end{lstlisting}

NTRU is a public-key encryption and key-encapsulation mechanism, which allows to safely transfer a session key between two (or more) parties over an insecure medium. NTRU is one of the four remaining finalists of NIST post-quantum project. 
A specific constant-time sorting sub-routine used in NTRU~\cite{ntrup} and NTRU Prime~\cite{ntrup}, is vulnerable to a power-based side-channel attack~\cite{2021:HOST}. 
A significant power consumption difference between the two possible outputs 0 and -1 during the execution of the targeted shift operation at line 6 in Listing~\ref{lst:ntru}. This vulnerability occurs due to the significant difference in the Hamming weight representations. The -1 ($0xFFFFFFFF$) shows a power consumption behavior for $\omega = 32$, while the output 0 ($0x00000000$) behavior matches with $\omega = 0$.

\end{document}